\documentclass{aa}
\usepackage[utf8]{inputenc}
\usepackage{graphicx}
\usepackage[english]{babel}
\usepackage{amsmath}
\usepackage{mathtools}
\usepackage{amssymb}
\usepackage{txfonts}
\usepackage{hyperref} 
\usepackage{color}
\usepackage{rotating}
\usepackage{enumitem}
\usepackage[labelfont=bf]{caption}
\usepackage{subcaption}
\usepackage{ragged2e}
\usepackage{threeparttable}
\usepackage{placeins}
\usepackage{url}
\hypersetup{colorlinks=true,citecolor=blue}

\definecolor{orange}{rgb}{0.99,0.69,0.07}

\newcommand{\fig}[1]{Fig.\,\ref{#1}}

\newcommand{\figss}[3]{Figs.\,\ref{#1},\,\ref{#2} and \ref{#3}}
\newcommand{\sect}[1]{Sect.\,\ref{#1}}

\newcommand{\app}[1]{Appendix\,\ref{#1}}
\newcommand{\tab}[1]{Table\,\ref{#1}}

\begin{document}

\title{First exploration of the runaway greenhouse transition with a GCM}
\author{Guillaume Chaverot\inst{1,2}, Emeline Bolmont\inst{1,2}, Martin Turbet\inst{3,4,1}}

\offprints{G. Chaverot,\\ email: guillaume.chaverot@unige.ch}

\institute{
$^1$ Observatoire astronomique de l'Universit\'e de Gen\`eve, chemin Pegasi 51, CH-1290 Versoix, Switzerland\\
$^2$ Life in the Universe Center, Geneva, Switzerland\\
$^3$ Laboratoire de M\'et\'eorologie Dynamique/IPSL, CNRS, Sorbonne Universit\'e, \'Ecole Normale Sup\'erieure, PSL Research University, \'Ecole Polytechnique, 75005 Paris, France\\
$^4$ Laboratoire d'astrophysique de Bordeaux, Univ. Bordeaux, CNRS, B18N, allée Geoffroy Saint-Hilaire, 33615 Pessac, France}

  \date{Accepted in A\&A}
  \abstract
   {Even if their detection is for now challenging, observation of small terrestrial planets will be easier in a near future thanks to continuous improvements of detection and characterisation instruments. In this quest, climate modeling is a key step to understand their characteristics, atmospheric composition and possible history. 
   
   If a surface water reservoir is present on such a terrestrial planet, an increase in insolation may lead to a dramatic positive feedback induced by water evaporation: the runaway greenhouse.
   The resulting rise of global surface temperature leads to the evaporation of the entire water reservoir, separating two very different population of planets: 1) temperate planets with a surface water ocean and 2) hot planets with a puffed atmosphere dominated by water vapor. 
   Therefore, the understanding of the runaway greenhouse is pivotal to assess the different evolution of Venus and the Earth, as well as every similar terrestrial exoplanet.
   
   In this work we use a 3D General Circulation Model (GCM), the Generic-PCM, to study the runaway greenhouse transition, linking temperate and post-runaway states. 
   Our simulations are made of two steps. First, assuming initially a liquid surface ocean, an evaporation phase which enriches the atmosphere in water vapor. Second, when the ocean is considered entirely evaporated, a dry transition phase for which the surface temperature increases dramatically. Finally, it converges on a hot and stable post-runaway state.
   By describing in detail the evolution of the climate during these two steps, we show a rapid transition of the cloud coverage and of the wind circulation from the troposphere to the stratosphere. By comparing our result to previous studies using 1D models, we discuss the effect of intrinsically 3D processes such as the global dynamics and the clouds, keys to understand the runaway greenhouse. 
   We also explore the potential reversibility of the runaway greenhouse, limited by its radiative unbalance.}
  

\titlerunning{First exploration of the runaway greenhouse transition with a Global Climate Model}
\authorrunning{Chaverot, Turbet, Bolmont} 

   \maketitle

\section{Introduction}

Continuous improvement of telescopes, instruments and techniques allows us to detect and characterize a growing number of exoplanets.
The new generation of instruments will enable to characterize atmospheres of small, rocky and temperate planets, which is now within reach thanks to the James Webb Space Telescope (JWST) \citep[e.g.][]{morley_observing_2017,lustig-yaeger_detectability_2019, fauchez_impact_2019, wunderlich_detectability_2019}.
The first results of the JWST will likely open a new era in the characterisation of such targets. Future astronomical ground-based instruments in development such as RISTRETTO@VLT \citep{lovis_ristretto_2022} or ANDES@E-ELT \citep{marconi_andes_2022} will allow to go beyond and study Earth-like planets. 
The possibility of detecting objects similar to our planet brings up the ancestral question of life (and ``habitability'') in the Universe.
Astrobiology is the field of science that aims to answer this question while exoplanetology is now a big piece in that gigantic puzzle.
In particular, an essential condition for the emergence and the maintenance of life as we know it, is the presence of liquid water. Therefore, modeling the potential climate of planets we will characterize is essential to determine the conditions under which water can exist in liquid phase at their surface or in the atmosphere as clouds. This modeling effort coupled with advanced spectroscopy techniques will allow us to make progress in the search for life in the Universe. 

For an Earth-like planet with a given amount of water, there is a range of distance to the host star for which this water can be liquid at the surface. Too close, the water evaporates and too far, it freezes. This ideal zone is the Habitable Zone (HZ) \citep{kasting_habitable_1993, kopparapu_habitable_2013}. 
This concept is extremely useful in the exoplanet community to statistically study large samples of targets under the spectrum of habitability and search for life.
The HZ is function of the brightness of the star, the size of the planet and the composition of its atmosphere \cite{kopparapu_habitable_2013, kopparapu_habitable_2014, kopparapu_habitable_2017}.
If the planet is close enough to the star, the temperature rises and oceans start to evaporate which may add enough vapor in the bottom part of the atmosphere to completely absorb the thermal emission because of the strong greenhouse effect of water vapor.
Once thermal emission reaches the Simpson-Nakajima limit \citep{simpson_studies_1929, nakajima_study_1992}, the planet is not able to cool down anymore and the positive feedback of the runaway greenhouse arises \citep{komabayasi_discrete_1967,ingersoll_runaway_1969, kasting_runaway_1988, goldblatt_runaway_2012}. 
This leads to the evaporation of the entire surface ocean and a dramatic increase in global surface temperature of several hundreds of Kelvin. This climatically unstable transition separates two population of planets: temperate planets and hot post-runaway planets \citep[e.g.][]{hamano_emergence_2013,turbet_day-night_2021}. This is one of the scenario to explain the difference between the Earth and early-Venus \citep{way_was_2016}.
Temperate planets are similar to the Earth with surface condensed water and a thin atmosphere, while hot planets have an extended water-dominated atmosphere with high surface temperature, up to a few thousand Kelvins \citep{turbet_runaway_2019}. 
The inner edge of the HZ is determined in part by the runaway greenhouse process.
The key question of the existence of surface liquid water on terrestrial planets is thus inherently connected to the runaway greenhouse with strong implications for our understanding of the possible evolution of exoplanets. 
As shown in \cite{turbet_revised_2020}, this process may also strongly influence the mass-radius relationships of terrestrial planets by clearly separating temperate planets hosting a water ocean and hot steam planets.

Regarding the importance of this process, many papers studied it using simple 1D climate models. 
But as shown by some 3D General Circulation Models (GCMs) studies \citep[e.g.][]{leconte_increased_2013,kopparapu_habitable_2017}, approximations assumed in 1D models lead to a wrong estimation of the climate. 
For example, Hadley cells on Earth-like planets tend to reduce the relative humidity around the tropics allowing a larger Outgoing Longwave Radiation (OLR) than the one estimated by 1D models \citep{leconte_increased_2013}. 
Moreover, specific clouds pattern may appear for slow rotating planets planets \citep{yang_stabilizing_2013} or for post-runaway states \citep{turbet_day-night_2021} which by definition cannot be accounted for in a 1D model.

Even if 1D models are useful in many cases because of their adaptability and fast computation time, more complex 3D models should be preferred to capture intrinsically 3D physical processes, such as clouds and global dynamics. 
Several studies explored the runaway greenhouse effect on Earth-like planets with different GCMs \cite[e.g.][]{leconte_increased_2013,yang_stabilizing_2013,wolf_delayed_2014,wolf_evolution_2015,popp_transition_2016, kopparapu_habitable_2017} in order to determine under which insolation this positive feedback arises. 
Unfortunately, highly unstable climates and high water vapor content may cause numerical instabilities or reveal model limitations. As explained by \cite{kopparapu_inner_2016} and \cite{kane_climate_2018}, usually, existing studies assume a runaway greenhouse when the simulation fails due to high water content and large differences between the OLR and the Incident Stellar Radiation (ISR). 
Some other studies found moist stables states \cite[e.g.][]{popp_transition_2016}, for which the planet can be in a stable regime with a wet troposphere and a global temperature lower than 350\,K. This prevent any dramatic huge increase of the global temperature leading to a post-runaway state, thus questioning the definition of the inner edge of the HZ as well as the implications for habitability.

The originality of our work is to push the study beyond the tipping point and to describe the evolution of the climate across the runaway greenhouse transition in order to link temperate and post-runaway states. We also explore the reversibility of the runaway greenhouse and we discuss the possible existence of an intermediate moist greenhouse state.
We describe in detail inherently 3D processes in order to have a complete and accurate overview of the evolution of the atmosphere, mainly through its enrichment in water vapor and the consequences of the modification of the cloud coverage. 
In \sect{sect_method} we describe the method followed to explore the runaway greenhouse transition, as well as the GCM we use and the considered numerical setups. 
We present our results in \sect{sect_results} and we discuss the limits and the prospects of this work in \sect{sect_discussion}. 

\section{Modeling the climate}
\label{sect_method}

\begin{table*}\footnotesize\centering
\setlength{\doublerulesep}{\arrayrulewidth}
\captionsetup{justification=justified}
\caption{Parameters of the simulations}
\label{table_param}
\begin{tabular}[c]{cc}
    \hline\hline\hline
    \multicolumn{2}{c}{\textbf{Planetary parameters}} \\
    \hline   
    Stellar spectrum & Sun \\
    Orbital period &  365\,days\\
    Rotation period & 24\,hours\\
    Obliquity & 23\textdegree\\
    Planetary radius & 6400\,km \\
    Gravity & 9.81 \\
    \hline\hline\hline
\end{tabular}\normalsize
\vspace{0.3cm}
\begin{tabular}[c]{ccccccc}
    \hline\hline\hline
    \textbf{Simulation} & W01 & W1 & W10 & WCO2 & E1 & ECO2 \\
    \hline
    Topography & waterworld & waterworld & waterworld & waterworld & Earth's continents & Earth's continents  \\
    Surface albedo & 0.07 & 0.07 & 0.07 & 0.07 & Earth's albedo & Earth's albedo   \\
    \hline
    H$_2$O & variable & variable & variable & variable & variable & variable \\
    N$_2$ & 0.1\,bar & 1\,bar & 10\,bar & 1\,bar & 1\,bar & 1\,bar \\
    CO$_2$ & - & - & - & 376\,ppm & - & 376\,ppm \\
    \hline\hline\hline
\end{tabular}\normalsize
\end{table*}

\subsection{Modeling the runaway greenhouse}
\label{sect_methodology}
Our method is similar to the one used by previous works \citep{yang_stabilizing_2013,leconte_increased_2013, wolf_delayed_2014,wolf_evolution_2015,popp_transition_2016,kopparapu_habitable_2017} which studied the onset of the runaway greenhouse with GCMs. 
First, we model a temperate planet similar to the present-day Earth then we increase step by step the insolation (ISR) to find the tipping point for which the well-known positive feedback arises. 
We keep the shape of the stellar spectrum unchanged when we modify the insolation.
The originality of this work is that we model the climate beyond the onset of the runaway greenhouse, that is for higher temperature and water vapor mixing ratio.
When the insolation is high enough to initiate a runaway greenhouse, we fix it to let the ocean evaporate in order to study the evolution of the climate during this critical transition phase.
The insolation and the global surface temperature for which this threshold arises depend on the configuration of the simulation (e.g. presence of continents, CO$_2$ content). 
Our aim is not here to find the exact value of the insolation corresponding to the tipping point, but to study the differences in the climate patterns.
It could be interesting to find the exact tipping point as done by \cite{leconte_increased_2013} or \cite{kopparapu_habitable_2017} for example, for various atmospheric compositions, but this is not the aim of this study. 
We perform simulations following this method for different atmospheric configurations: with/without Earth's continents, with/without CO$_2$ and for different N$_2$ pressures (see \tab{table_param}). 
In this article, our reference simulation is the waterworld including 1\,bar of N$_2$, without CO$_2$ (W1), as a representative example of the evolution of the climate during the runaway transition. Other simulations presented in \tab{table_param} are discussed in the context of a sensitivity study.

The key process allowing a runaway greenhouse is the evaporation of the surface ocean. 
To model the ocean, we use a two-layer slab ocean without heat transport, coupled to the atmosphere (see \sect{sect_setup}). 
Evaporating the totality of the Earth's oceans will induce a vapor surface pressure of 273\,bar. Regarding the large heat capacity of the water vapor, any temperature modification will be extremely slow in this situation, making such simulation unachievable from a numerical point of view.
Moreover, the timescale required to evaporate enough water to reach this pressure are also hugely long. 
To circumvent that difficulty, we choose to stop the evaporation process when a given quantity of water is evaporated (see evaporation phase described in \sect{sub_evap}). For numerical reasons, we test only two different evaporated water vapor pressures: 1 and 1.5 bar which correspond respectively to 10 and 15m of global layer equivalent (GEL), that is the depth of the equivalent liquid ocean covering the entire planet.
Then we remove the ocean from the simulation (i.e. we turn off the surface evaporation and we consistently adjust the surface albedo and the thermal inertia) and we set a new simulation by using the final state of the evaporation phase as the initial state (this is the dry transition phase described in \sect{sub_dry}). This method allows us to study the impact of the size of the water reservoir on the obtained post-runaway state.

Based on the conclusions of \cite{chaverot_how_2022} on the radiative transfer for non dilute water atmospheres, we use opacity data taking into account the transition of broadening species from nitrogen dominated to water dominated. 

\subsection{Numerical setup}
\label{sect_setup}
The Generic-PCM (G-PCM) \citep[e.g.][]{wordsworth_gliese_2011, leconte_increased_2013, turbet_habitability_2016, turbet_modeling_2018,fauchez_impact_2019, turbet_day-night_2021}, previously known as the LMD Generic GCM, has been adapted to model water-rich atmospheres \citep{leconte_increased_2013}. This makes it particularly appropriate to study the runaway greenhouse. 
The horizontal resolution is 64$\times$48 (longitude $\times$ latitude) with 30 vertical levels from 0.1, 1 or 10\,bar (depending on the simulation setup, see \tab{table_param}) to 10\,Pa. 
We consider the Earth's obliquity and all the simulations are initialized assuming summer in the northern hemisphere.
The dynamical time step is equal to 90\,seconds. The physics (evaporation, convection, etc. ) is computed every 15\,minutes while the radiative transfer is computed every hour.
The radiative transfer, assuming a Sun-like star, is done using a correlated-k table from \cite{chaverot_how_2022} for the mixture H$_2$O+N$_2$ and from \cite{leconte_increased_2013} for the mixture including 376ppm of CO$_2$.
The absorption line-list are taken from HITRAN 2012 \citep{rothman_hitran2012_2013} and HITRAN 2016 \citep{gordon_hitran2016_2017} respectively. 
We include also the N$_2$–N$_2$ collision-induced absorption (CIA) from the HITRAN CIA database \citep{karman_update_2019}.
The H$_2$O-H$_2$O and H$_2$O-air continua are from the version 3.3 of the MT\_CKD database \citep{mlawer_development_2012}.  
In our simulation, water is a variable gas which may condense or evaporate from the surface ocean but also in the atmosphere, and the relative humidity is computed consistently. The subgrid-scale dynamical processes (turbulent mixing and convection) as well as the moist convection scheme are described in \cite{leconte_increased_2013}.
The number of cloud condensation nuclei (CCNs) per unit mass of moist air used is fixed: $4\times10^6$~kg$^{-1}$ for liquid water clouds and $2\times10^4$~kg$^{-1}$ for ice clouds, while their radius is variable (see \citealt{leconte_increased_2013} for more details).

The model includes also a 2-layers slab ocean without heat transport \citep{codron_ekman_2012, charnay_exploring_2013}. 
The thermal inertia is equal to 18,000~J.s$^{-1/2}$.m$^{-2}$.K$^{-1}$ for oceans and 2,000~J.s$^{-1/2}$.m$^{-2}$.K$^{-1}$ for continents. The albedo of the ocean is 0.07 and varies for continents up to 0.35 following \cite{leconte_increased_2013} 
The heat capacity ($c_p$) and the mean molecular weight are consistently initialized for every setup but they stay constant during the simulation which means that we do not account for the modification induced by the addition of water vapor. 

\section{The runaway greenhouse transition}
\label{sect_results}

In this section we present evolutions of the climate from temperate stable states to stable post-runaway states, following the method described in \sect{sect_method}. 
First, we describe the onset of the runaway greenhouse (\sect{sub_evap}), here named \textit{evaporation phase} in reference to the enrichment of the atmosphere in water vapor. 
Second, we describe the \textit{dry transition phase} (\sect{sub_dry}) for which the ocean is considered entirely evaporated.

\subsection{Onset of the runaway greenhouse}
\label{sub_evap}

\subsubsection{Evolution of the OLR during the onset}
\label{sub_OLR}
We performed simulations for the different configurations presented in \tab{table_param} in order to quantify the influence of continents and strong infrared (IR) absorbers like CO$_2$ on the onset of the runaway greenhouse.

\begin{figure}[!ht]
    \centering\includegraphics[width=\linewidth]{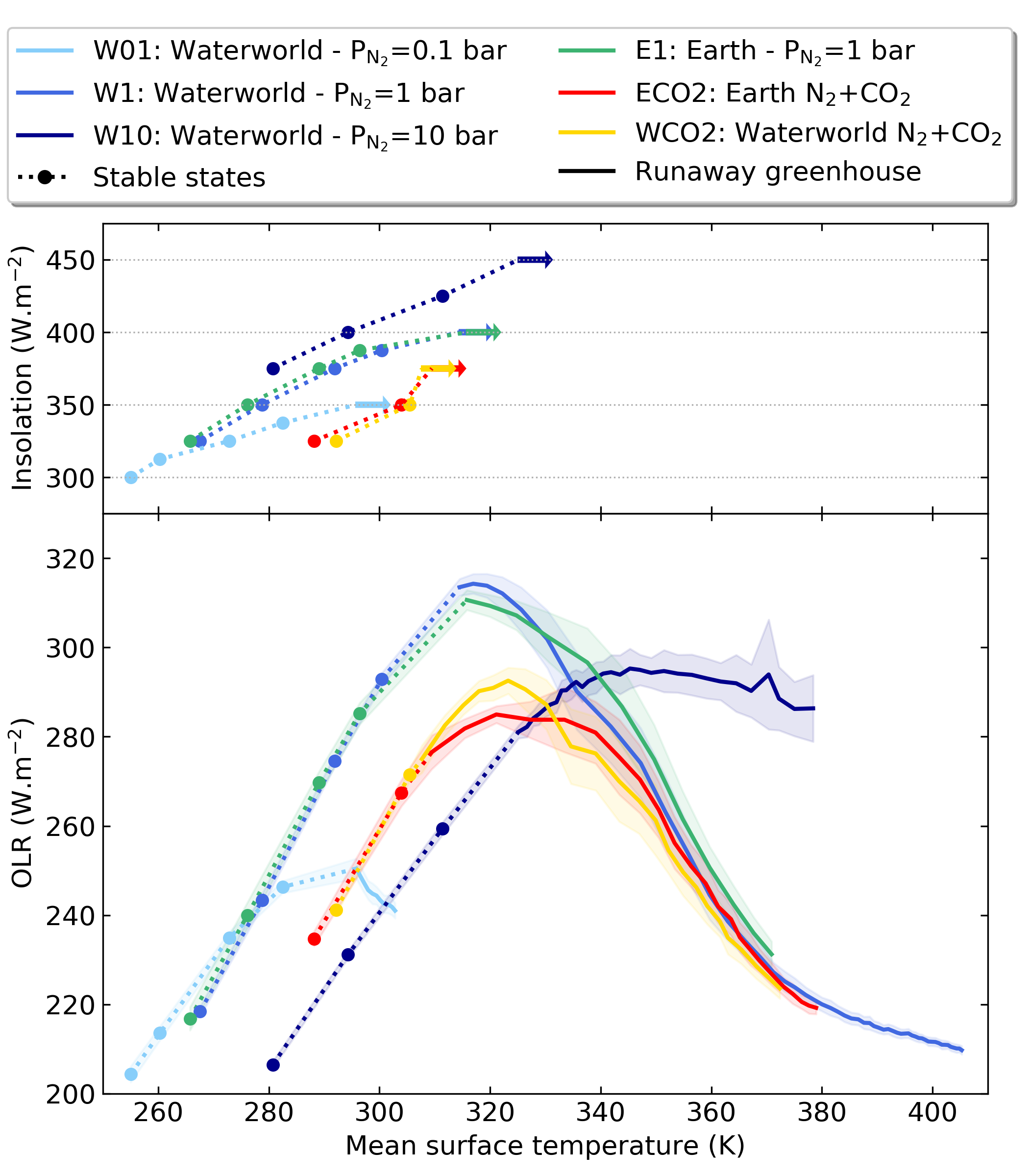}
    \caption{The bottom panel gives the value of the OLR function of the global surface temperature for each simulation setup described in \tab{table_param}. Colored dots are the stables states corresponding to various insolations while the full curves are the instable state of the runaway greenhouse. The top panel indicate the insolation values of each stable state. The colored arrows represent the insolation for which the runaway greenhouse arises for each setup. The OLR and temperature values are averaged over 2 years (for both the stable states and the runaway greenhouse) and the colored fills are the 1-sigma uncertainties due to variability on the OLR calculation.}
    \label{fig_OLRvsT}
\end{figure}

Figure \ref{fig_OLRvsT} provides an overview of the different simulations done. 
The bottom panel is the OLR as a function of the mean surface temperature while the top panel indicates the different values of the insolation considered for every setup. 
in \fig{fig_OLRvsT}, stable states, in other words simulations for which an equilibrium is possible, are indicated using black dots. 
The surface temperature of stable states is function of the insolation, therefore we find different stables states by increasing step-by-step the insolation (colored dots on top panel of \fig{fig_OLRvsT}) as described in \sect{sect_methodology}. 
When the insolation is high enough, a tipping point is reached and the atmosphere enters an unstable state (colored arrows on top panel of \fig{fig_OLRvsT}). Due to a large amount of water vapor, the bottom part of the atmosphere becomes optically thick to the thermal emission and the OLR is limited.
Therefore, there is no longer equality between the OLR and the Absorbed Stellar Radiation (ASR) (see \fig{fig_OLR_OLRcsvsT}) and the global temperature increases continuously due to this radiative imbalance (solid colored lines on the bottom panel of \fig{fig_OLRvsT}).

For every configuration presented in \fig{fig_OLRvsT} the relation between the OLR and the mean surface temperature for stable states is roughly linear which is consistent with previous studies \cite[e.g.][]{leconte_increased_2013, wolf_evolution_2015}.
For a given insolation, continents tend to reduce the global surface temperature of few Kelvins because of their high albedo (higher than the deep ocean) as shown on the top panel of \fig{fig_OLRvsT} (W1 vs E1 and WCO2 vs ECO2). 
For atmospheres with 1\,bar of nitrogen (W1, E1, ECO2 and WCO2), the evolution of the OLR during the runaway transition (i.e. when the surface temperature reach $\sim$340~K) is not sensitive to the presence of continents and/or the presence of low concentration of CO$_2$ for a given total surface pressure: every simulation converge roughly on a same OLR curve.  
This is because when the surface temperatures are high enough, background gases become negligible compared to the radiative effect of high vapor pressures \citep{chaverot_how_2022}. 
Moreover, the radiatively active part of the atmosphere is decoupled from the surface conditions, therefore the surface albedo (i.e. continents) does not influence the radiative balance anymore.
For higher background gas pressures (W10 in \fig{fig_OLRvsT}) the onset of the runaway greenhouse arises at higher temperature and for a higher insolation (450\,W.m$^{-2}$). 
For this extreme case with 10\,bar of nitrogen, the temperature profile is closer to a purely dry adiabatic profile which is "less steep" than a wet adiabatic profile. Consequently, for a given global surface temperature the atmosphere is colder thus dryer. Therefore, even if there is less water in the atmosphere (vapor and clouds) the averaged OLR is lower as shown by \fig{fig_OLRvsT}.
This effect of high nitrogen pressures on the steepness of the temperature profile is also discussed in \cite{chaverot_how_2022}, but they do not find smaller OLR values for high nitrogen pressure. This is probably linked to the pseudo-adiabatic assumption fixing the water vapor partial pressure as a function of the temperature. This assumption usually done in 1D models fixes the relative humidity to unity and underestimates the OLR \citep{ishiwatari_numerical_2002,leconte_increased_2013}.
On the other side, lower background gas pressures (0.1\,bar of nitrogen, see W01 in \fig{fig_OLRvsT}) lead to a runaway greenhouse transition at a lower temperature, OLR value and insolation (350\,W.m$^{-2}$). A lower OLR bump for a low nitrogen pressure is consistent with results obtained by \cite{chaverot_how_2022} with a 1D model. The strong pressure broadening of H$_2$O-H$_2$O collisions makes the absorption more efficient than in the diluted case assuming 1\,bar of nitrogen. The temperature profile is also closer to a wet adiabat which warms the stratosphere (compare to the dry adiabat case) and trigger the onset of the runaway greenhouse.

Finally, adding 376~ppm of CO$_2$ shifts the onset of the runaway greenhouse to lower insolation (375\,W.m$^{-2}$ for ECO2 and WCO2 instead of 400~W.m$^2$ for E1 and W1) because of the strong greenhouse power of CO$_2$. 
This drop is discussed using simpler 1D models as well \citep[e.g.][]{nakajima_study_1992,goldblatt_runaway_2012,goldblatt_low_2013, koll_hot_2019,chaverot_how_2022}.
For a given insolation the global surface temperature is also about 20\,K higher in presence of CO$_2$ (top panel in \fig{fig_OLRvsT}) which is consistent regarding the greenhouse power of this gas.

During the runaway greenhouse, the insolation is kept unchanged to let the system evolve. The OLR reaches a maximum then decreases strongly. This is due to a combination of radiative effects and a quick modification of the cloud pattern. 
Spectroscopic processes are described by using a 1D radiative-convective model in \citet{chaverot_how_2022} and the modification of the cloud coverage is described in \sect{sub_cloud_coverage}. 
In \fig{fig_OLRvsT}, the colored envelope delimiting the one-sigma variability of the OLR increases largely during the first steps of the runaway greenhouse. 
The continuous evaporation characteristic of this process densifies the cloud coverage. 
For medium temperatures (between 320~K and 360~K), the global dynamics induces variable patterns in this cloud coverage, thus strong variations of the spatial distribution of the OLR. 
When the global surface temperature is high enough to maintain a large amount of water vapor in the atmosphere (beyond 360~K), the cloud coverage becomes more homogeneous reducing the thermal emission windows thus the OLR as well as its variability. 

\subsubsection{Radiative imbalance}
\label{sub_rad_imbalance}

\begin{figure}[!ht]
    \centering\includegraphics[width=\linewidth]{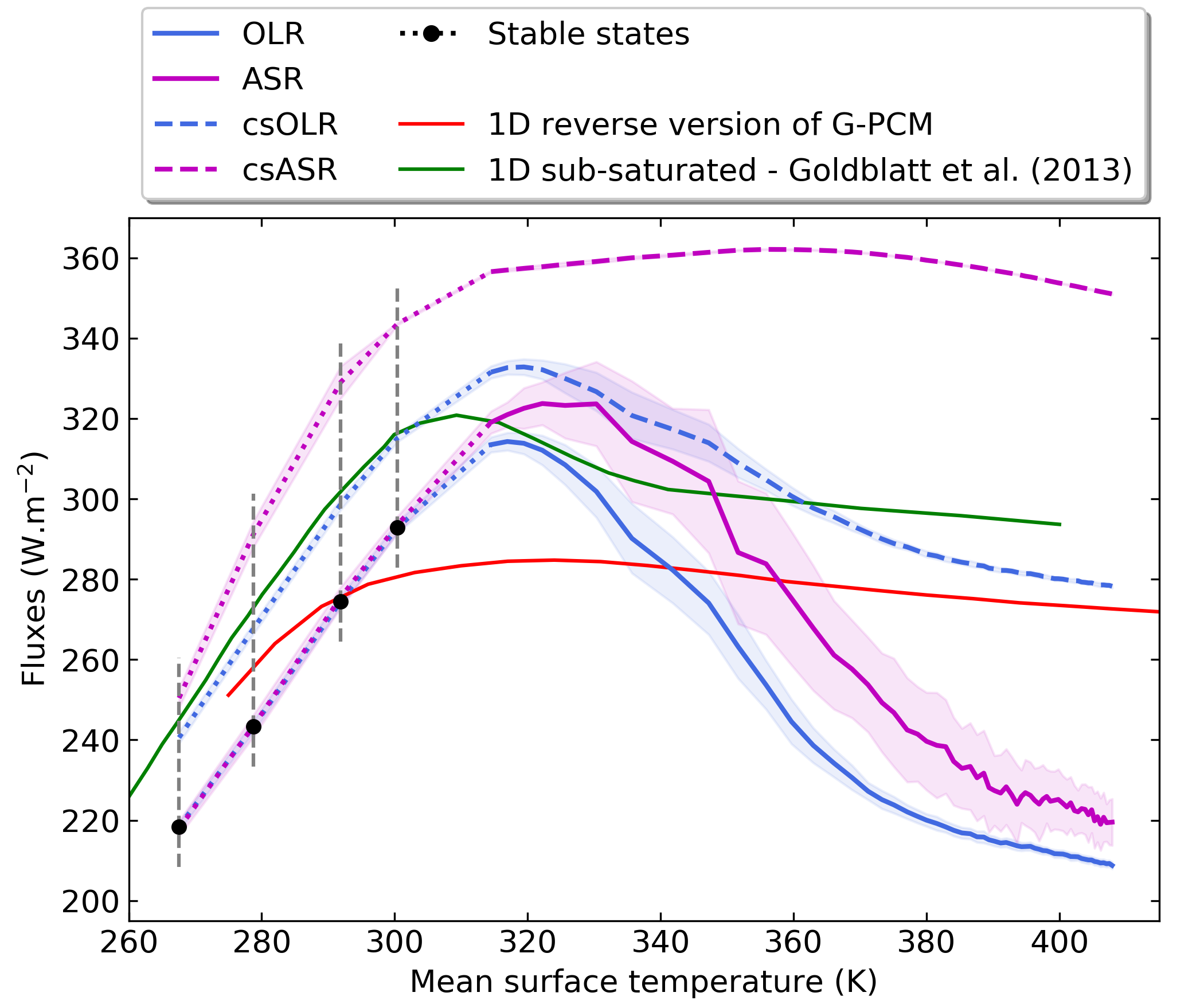}
    \caption{Thermal emission and absorbed flux as a function of the mean surface temperature for a waterworld with 1 bar of nitrogen without CO$_2$ (W1). Blue and purple solid lines are OLR and ASR respectively while blue and purple dashed lines are their clear-sky equivalent (csOLR and csASR). Black dots are the stable states and the blue solid line is the runaway greenhouse for an insolation equal to 400\,W/m${^2}$. For comparison the red line is the OLR computed with the 1D reverse version of the Generic PCM which uses the same radiative transfer module and assume a fully saturated atmosphere (named 1D reverse version of LMD Generic in \citealt{chaverot_how_2022}). The green line is the OLR computed using a 1D reverse model assuming a subsaturated atmosphere (saturation mixing ratio less than 5\%) from \cite{goldblatt_low_2013}. The flux and temperature values are averaged over 2 years and the colored areas are the 1-sigma uncertainties due to variability on the fluxes calculation.}
    \label{fig_OLR_OLRcsvsT}
\end{figure}

To study the evolution of the cloud coverage and its radiative effects, we compare in \fig{fig_OLR_OLRcsvsT} the OLR and the ASR to their clear sky equivalent (csOLR and csASR). To compute csOLR and csASR, we run a simulation including clouds but we neglect their radiative effect to estimate the emitted and absorbed fluxes. 
On the same figure, the red curve correspond to OLR values obtained using the 1D cloud-free version of the Generic PCM\footnote{named 1D reverse version of LMD Generic in \cite{chaverot_how_2022}} described in \cite{chaverot_how_2022}, with an atmosphere including H$_2$O and 1\,bar of N$_2$ and a relative humidity value fixed to unity. The aim is to compare 1D and 3D models sharing the same radiative transfer calculation.

At low and high temperatures, respectively when the atmosphere is nitrogen or water dominated, csOLR values from 1D and 3D simulations tend to converge. 
As there is no clouds in the 1D simulation, the only remaining difference is the absence of dynamics in the 1D model. Therefore, at low temperature the atmosphere is very thin and the csOLR is not much affected by the global dynamics. In other word, it is close to the Planck function. 
On the other side, at high temperature, the water vapor is well distributed around the planet and the csOLR is not affected by the dynamics. 
For intermediate temperatures, and especially for the onset of the runaway greenhouse when the evaporation begins, the distribution of the vapor is not homogeneous and the 1D version of the Generic PCM largely underestimates the global OLR.
As explained in \cite{ishiwatari_numerical_2002} and \cite{leconte_increased_2013}, this underestimation of the OLR of 1D models assuming a saturated atmosphere is due to the downward branches of the Hadley cells which reduce the relative humidity around the tropics, creating radiative windows by reducing the quantity of water vapor.
To circumvent this issue, 1D models with variable humidity should be preferred \citep[e.g.][]{ramirez_can_2014}.
In \fig{fig_OLR_OLRcsvsT}, the green curve which is from \cite{goldblatt_low_2013} is obtained by using a cloud free 1D model assuming a saturation mixing ratio of less than 5\%. This allows to fit accurately csOLR from GCM simulations for temperate states. Even if some differences remain for the runaway greenhouse, the shape of csOLR and particularly the bump is quite well represented. The major difference between results from 1D and 3D simulations is thus induced by the evolution of the cloud coverage. The differences remaining between \cite{goldblatt_low_2013} (green curve) and csOLR could come from dynamics, the different opacity database used or a mix of the two. Indeed, \cite{goldblatt_low_2013} use HITEMP while we use HITRAN in this work.

The difference between the OLR and csOLR curves in \fig{fig_OLR_OLRcsvsT} is only due to the radiative effect of the clouds. While this difference is rather constant for stable states, it grows during the runaway greenhouse. 
This is due to the strong evaporation which allows a densification and a drift of the cloud coverage toward the upper layers of the atmosphere (see \sect{sub_cloud_coverage}) increasing their radiative effect (see also \fig{fig_cloud_forcing} showing the usual cloud forcing metrics). This is visible through the sharp increase of the Bond albedo and the drop of the ASR (pink solid line in \fig{fig_OLR_OLRcsvsT}).
Nevertheless, as the OLR decreases as well, the effective flux (Seff=ASR/OLR) is greater than unity and the atmosphere warms-up. 
The energy imbalance is constant beyond 390\,K ($\approx$12\,W.m$^{-2}$) and never exceed $\approx$25\,W.m$^{-2}$ while it reaches $\approx$150\,W.m$^{-2}$ in \cite{kopparapu_habitable_2017} for a global surface temperature equal to 350\,K around a 3700\,K star. 
We found that this energy imbalance may reach higher values for higher insolations (e.g. $\approx$80\,W.m$^{-2}$ for ISR=475\,W.m$^{-2}$ instead of ISR=400\,W.m$^{-2}$ for the reference simulation W1).

During the runaway greenhouse, the difference between the csOLR and the csASR increases, highlighting the increase of the water vapor content in the atmosphere, and thus of its absorption. 
This means that the OLR drops due to a combined absorption induced by the clouds and by a efficient horizontal homogenization of the the increasing content of water vapor in the atmosphere. As the csASR is roughly constant during the runaway greenhouse, the drop of ASR is mainly due to the increase of the Bond albedo due to the evolution of the cloud coverage.

\subsubsection{Evolution the cloud coverage}
\label{sub_cloud_coverage}

\begin{figure}[!ht]
    \centering\includegraphics[width=\linewidth]{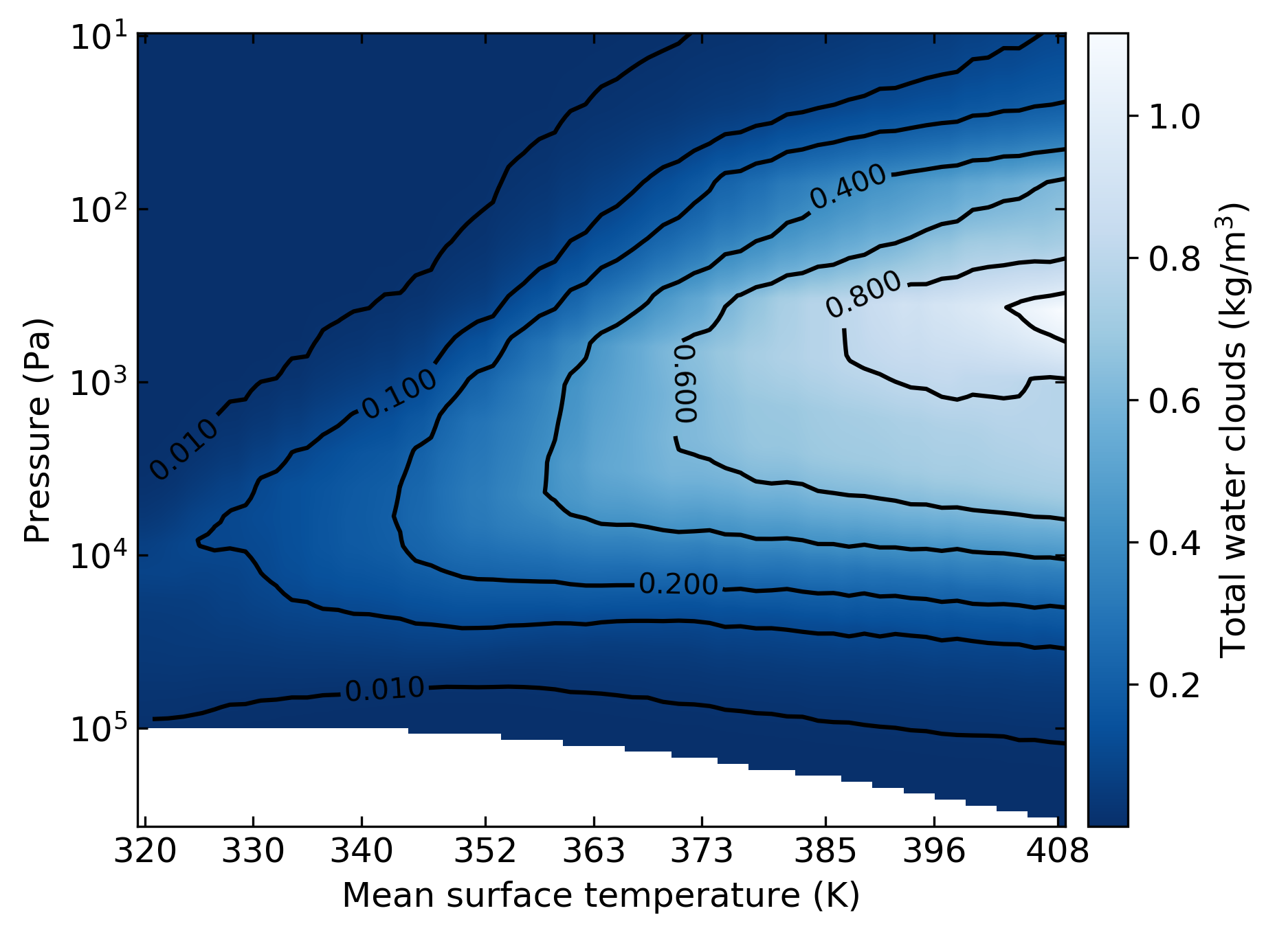}
    \caption{Map of the evolution of the mean vertical profile of the total water cloud (liquid + ice) content as a function of the global surface temperature during the evaporation phase. The simulation setup is the waterworld with 1 bar of nitrogen without CO$_2$ (W1). The profiles are averaged over one year. }
    \label{fig_Vmap_clouds}
\end{figure}

\begin{figure*}[!ht]
    \centering\includegraphics[width=\linewidth]{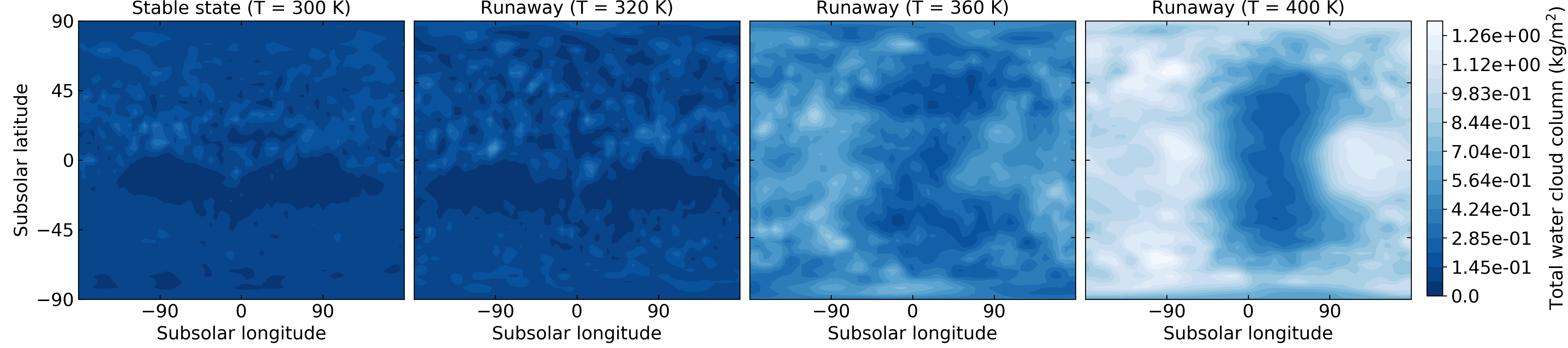}
    \caption{Latitude-longitude map of the clouds (liquid + ice) distribution during the evaporation phase. The maps were calculated in the heliocentric frame, that is keeping the subsolar point at 0° longitude and 0° latitude, and using an annual average. The simulation setup is the waterworld with 1 bar of nitrogen without CO$_2$ (W1).}
    \label{fig_latlon_clouds}
\end{figure*}

As explained in \sect{sub_rad_imbalance}, during the runaway greenhouse the thickness of the cloud coverage grows (due to a strong evaporation) and the altitude of the cloud deck drifts toward the upper layers of the atmosphere.
This effect is visible through the evolution of the mean vertical profile of the cloud coverage as a function of the temperature given by \fig{fig_Vmap_clouds}. The increase of surface pressure through the evaporation of the ocean is also visible on the same figure. The same evolution of the clouds has been observed by \cite{wolf_evolution_2015} using CAM4.
The water cloud content is constant in the bottom part of the atmosphere while the top-of-atmosphere layers are enriched from $\approx$10$^{-7}$ to $\approx$10$^{-1}$\,kg.m$^{-3}$.
In the same way, the stratosphere becomes wetter with the augmentation of temperature. At 400\,K the vapor profile is quasi-constant and tends toward 1\,kg.kg$^{-1}$ (see \app{app_vapor} for more details). This tendency is consistent with the results obtained by \cite{turbet_day-night_2021} studying the runaway greenhouse with hot and water-dominated initial conditions. 
As the quantity of clouds at the top of the atmosphere increases, the Bond albedo increases (from 0.2 to 0.45, see \fig{fig_albedoT_all}), inducing a drop of the ASR as discussed in \sect{sub_rad_imbalance}.

The day-to-night distribution of the clouds is not homogeneous during the runaway greenhouse, as shown on the latitude-longitude maps averaged over a year and centered at the substellar point presented in \fig{fig_latlon_clouds}. 
The pattern observed at 400\,K (right panel on the figure) is similar to that observed and discussed in \cite{turbet_day-night_2021} for hot and steam post-runaway atmospheres except that here a liquid surface ocean is still present, and thus the albedo on the day side is not strictly zero. 
The ocean allows cloud formation on the day-side, even if they are mainly located on the night-side.
The cloudy night-side provides an efficient greenhouse effect, contributing to the warming of the atmosphere.
While non-existent for temperate stable states, the contrast of this specific pattern amplifies during the evaporation of the ocean. 
In this work, we assume a rapidly rotating Earth-like planet which induces an uniform cloud coverage of temperate stables states. Previous studies \citep{yang_stabilizing_2013, kopparapu_habitable_2017} have shown that slow-rotating planets preferentially form cloud at the sub-stellar point for temperate stables states, preventing an intense warming and delaying the runaway greenhouse thanks to a high albedo on the day-side. 
These two different scenarios are discussed in \cite{turbet_day-night_2021}. 
The evaporation of the clouds at the sub-stellar point during the onset of the runaway greenhouse described in detail in this section was also observed by \cite{kopparapu_habitable_2017} for planets orbiting various stellar types.

\subsubsection{Atmospheric circulation}
\label{sub_winds}

\begin{figure*}[!ht]
    \centering\includegraphics[width=\linewidth]{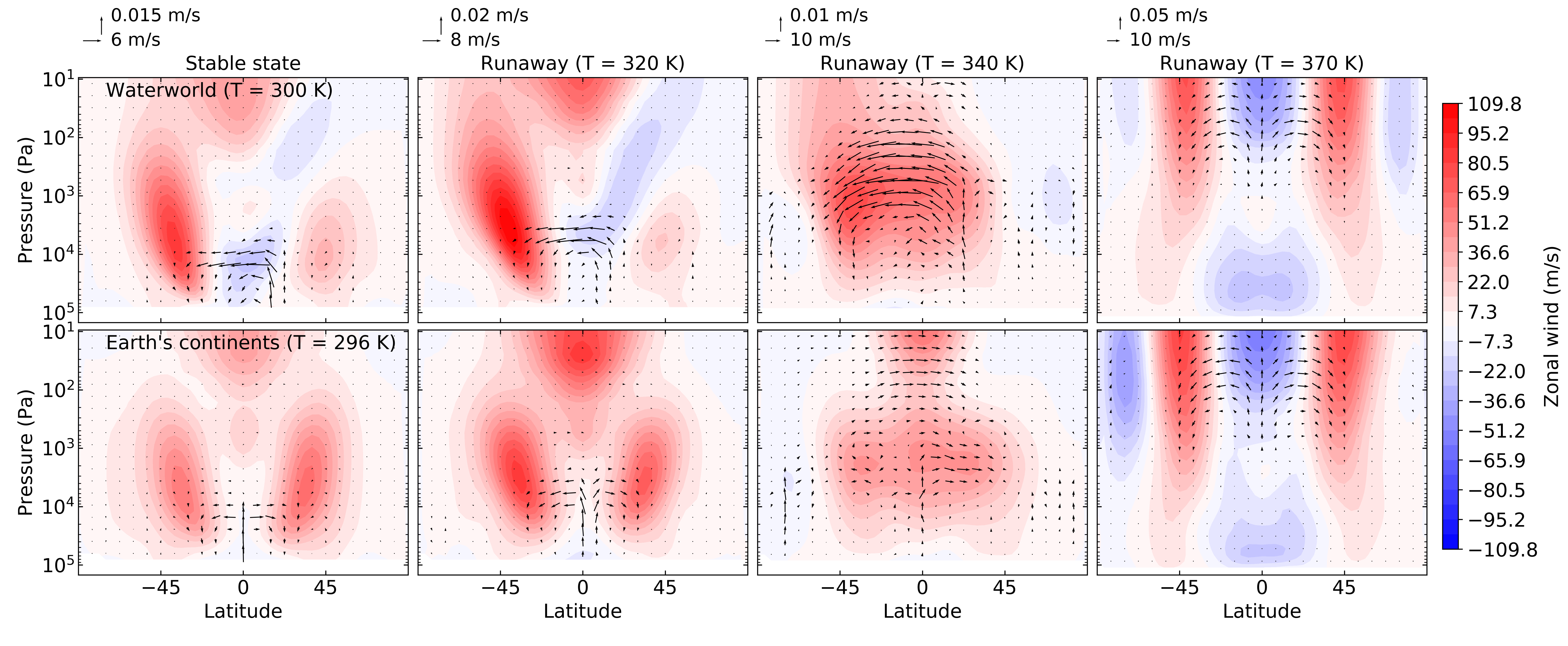}
    \caption{Evolution of the wind circulation during the runaway greenhouse. The color map represent the zonal wind (positive values represent eastward winds) and the arrows the vertical and meridional winds averaged over 1 year. 
    The first row is the evolution of the winds for the waterworld setup (W1) and the bottom row is considering the Earth's continents (E1), both without CO$_2$. The first panel of each row is the warmer stable state obtained for the two considered setups at I=375\,W.m$^2$. Others show the evolution during the runaway greenhouse for different temperatures. Wind speeds corresponding to the length of the arrows vary with temperature.
    }
    \label{fig_wind}
\end{figure*}

The global circulation of the atmosphere changes during the runaway greenhouse. 
The usual Hadley cells are visible in \fig{fig_wind} for temperate states (bottom left panel), with an upward branch at the equator and downward branches at the tropics. This terrestrial circulation disappears during the runaway greenhouse while a new circulation arises at the top of the atmosphere (bottom right). 
As the water vapor pressure increases in the bottom part of the atmosphere, the opacity increases and there is less and less stellar flux reaching the surface. 
Because of this, the temperature is more homogeneously distributed around the planet which decreases the surface wind jets. 
On the other side, as the absorption in the upper layers rises the radiative heating strongly increases, inducing strong meridional jets at the top of the atmosphere.
This stratospheric circulation characteristic of the runaway greenhouse and decorrelated from the surface conditions is visible in every simulation we performed. 
The same circulation was found by \cite{turbet_day-night_2021} for similar conditions and by \cite{fujii_nir-driven_2017}, using ROCKE-3D, for water-dominated tidally locked planets.

As described in \sect{sect_method} we use a slab ocean without heat transport. For simulations without continents we observe a hemispheric difference of the average annual temperature. More precisely, the northern hemisphere is warmer.
This hemispheric difference can also be seen in \fig{fig_latlon_clouds} where the cloud coverage is slightly more important for the two left panels.
This induces an atmospheric circulation around the equator along the North-South direction that is visible on the two first panels of \fig{fig_wind}.
In our simulations, the atmosphere is initially water vapor free and the surface temperature is homogeneous. 
Due to the non-zero obliquity of the planet (see \tab{table_param}), heating and intense evaporation happen alternatively in both hemispheres. 
For the waterworld setup, the evaporation of the ocean is continuous over time, unlike the Earth's continent setup for which the evaporation rate over time is perturbed by the presence of continents during the rotation of the planet. 
Consequently, for a given insolation, the atmosphere of the waterworld hosts more water vapor. 
In addition to this, the non-symmetric evaporation due to the obliquity of the planet induces a difference of the opacity between the two hemispheres. 
Therefore, this creates a North-South forcing of the heat transport, and one Hadley cell overrides the other.
The downward branch of the Hadley cell reduces the relative humidity as described in \cite{leconte_increased_2013} which allows locally a larger thermal emission, thus preventing the evaporation. This induces a feedback which maintain this specific circulation.
This modification of the Hadley cells is not function of the seasons and seems stable along the entire year.
When the runaway greenhouse occurs, as the physics is de-coupled from the surface this circulation weakens to converge on a unique stratospheric circulation pattern (last two panel in \fig{fig_wind}, see also \citealt{fujii_nir-driven_2017})

This non-symmetric temperate circulation seems stable in the simulation W1 as the zonal averaged temperature is constant over time as well as the water vapor pressure. However, as the numerical integration time is limited by our resources, a very long equilibrium time (longer than a hundred years) should not be excluded. Even if this state seems stable, it is function of the initial conditions of the simulation. The north to south hemispheric circulation shown in \fig{fig_wind} (top left panel) corresponds to a simulation initialized assuming summer in the northern hemisphere. By initializing the simulation assuming summer in the southern hemisphere, the obtained circulation is reversed, meaning that the dominant wind is from south to north.
For simulations with continents (E1 and ECO2), the evaporation rate and the cloud distribution are less homogeneous along the longitude than for their waterworld equivalents (W1, WCO2). This disturbs the feedback described above, thus allowing more symmetrical circulation around the equator that is similar to Hadley cells we see on Earth.  
Moreover, a symmetrical circulation is observed for a planet without obliquity as well as for low insolations (i.e. global surface temperatures below 280\,K). 
This confirms that the modified circulation can happen preferentially on a waterworld, and it is induced by combining obliquity and high evaporation rates due to an insolation higher than the one of the Earth. 
Nevertheless, the sensitivity of this particular circulation needs to be deeply investigated to understand the contribution of each processes on it, as well as its stability criteria.
Moreover, the oceanic circulation could largely affect it. 

\subsubsection{Reversibility of the runaway greenhouse}
\label{sub_reverse}

\begin{figure}[!ht]
    \centering\includegraphics[width=\linewidth]{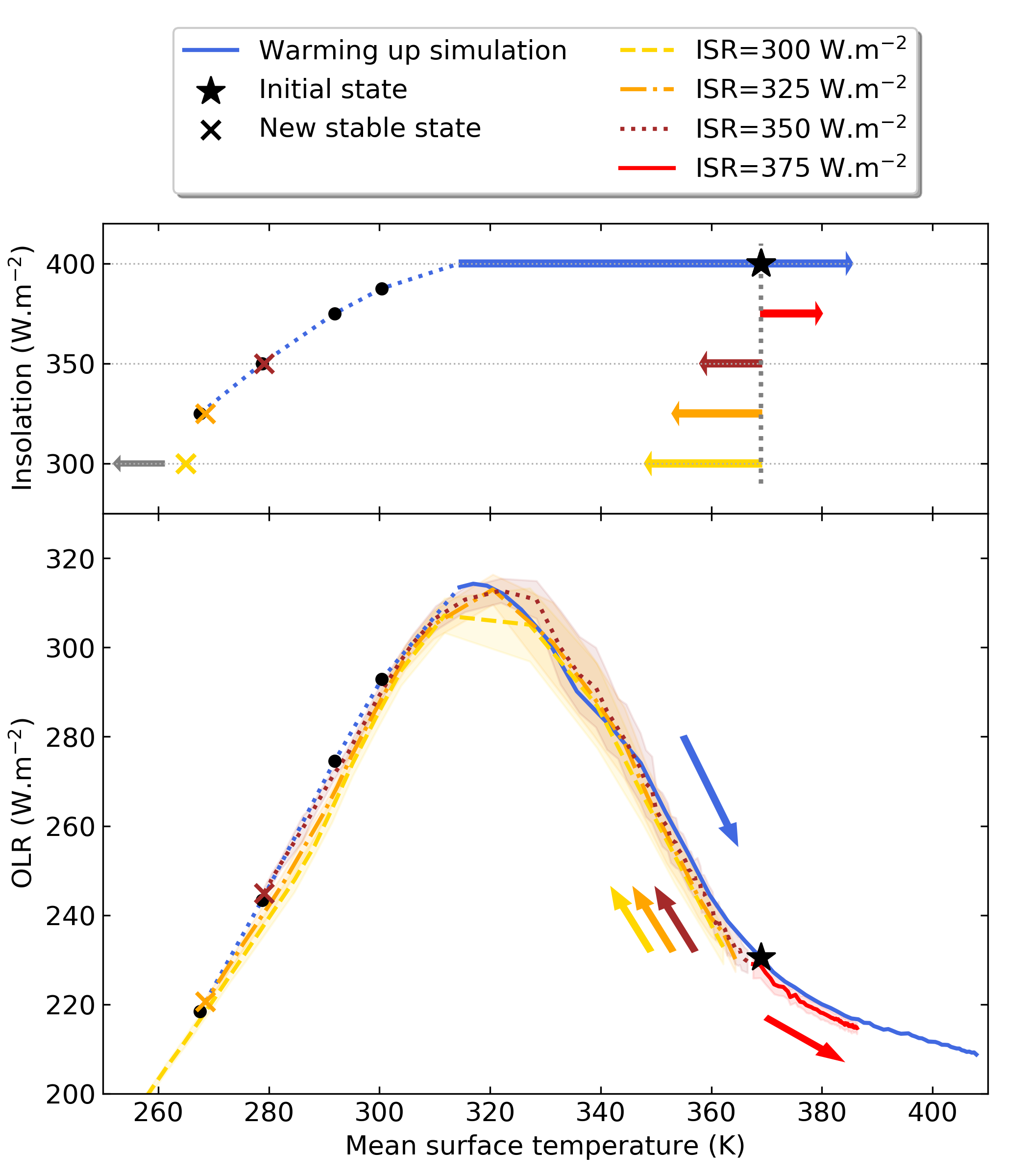}
    \caption{The bottom panel is the OLR function of the mean surface temperature. The top panel is the insolation of each simulation. The blue curve is the warming simulation presented in \fig{fig_OLRvsT} with an insolation equal to 400\,W.m$^{-2}$. Black dots are stable states assuming a 2-year average. Other colored curves (yellow, orange, brown and red) are the evolution of the OLR for lower insolations and assuming a initial temperature equal to 369\,K (the black star). Arrows are representing the tendency of the evolution of the global surface temperature (warming or cooling) for different insolations. Colored crosses represent the stable stable found for cooling simulations (ISR=300\,W.m$^{-2}$, 325\,W.m$^{-2}$ and 350\,W.m$^{-2}$ respectively). The top panel highlight the hysteresis loop induced by the clouds feedback during the runaway greenhouse. However, the bottom panel shows that there is no dependence of the OLR on the insolation for a given surface temperature. The simulation setup is a waterworld with 1 bar of nitrogen without CO$_2$.}
    \label{fig_OLR_reversed}
\end{figure}

In each simulation presented previously, we study the evolution of the atmosphere during the warming phase of the runaway greenhouse transition. 
We performed a set of simulations, presented in \fig{fig_OLR_reversed}, in which we start from a planet in runaway and reduce the insolation it receives in order to condense back the vapor into the surface ocean.
First, a simulation following the method described in \sect{sect_method} is performed in order to partially evaporate the ocean (blue curve in \fig{fig_OLR_reversed}) (see W1 in \tab{table_param}).
This induces an increase of the global surface temperature, symbolized by the blue arrow in \fig{fig_OLR_reversed}. 
Then we select a runaway greenhouse state with a global temperature equal to 369\,K as initial state for a second set of simulation (shown as a black star in \fig{fig_OLR_reversed}).
For this second step, we decrease the insolation compared to the one required to trigger a runaway greenhouse transition (ISR=400\,W.m$^{-2}$) in order to find the insolation required to condense back the vapor into the surface ocean (i.e. to cool down the planet).
The bottom panel of \fig{fig_OLR_reversed} presents the obtained OLR curves for different insolations (yellow: ISR=300\,W.m$^{-2}$, orange: ISR=325\,W.m$^{-2}$, brown: ISR=350\,W.m$^{-2}$, red: ISR=375\,W.m$^{-2}$). 
Colored arrows on the two panel represent the tendency evolution of the global surface temperature for each corresponding insolation (cooling or warming). 

The non-zero difference between the OLR and the ASR during the runaway (see \fig{fig_OLR_OLRcsvsT}) induces that a small decrease of insolation is not enough to condense back the water vapor and cool the planet. While ISR=400\,W.m$^{-2}$ is required to initiate the runaway greenhouse for the W1 setup, ISR=375\,W.m$^{-2}$ is not low enough to condense back the water vapor onto the surface starting at a global surface temperature of 369\,K. 
An insolation equal to 350\,W.m$^{-2}$  is able to cool down the planet, converging on a temperate stable state. For every cooling simulation, we recover the same stable stables as those found by increasing step-by-step the insolation (see \sect{sect_methodology}).
We show in \fig{fig_OLR_reversed} that during this cooling phase, the OLR increases following the same pathway than the warming W1 simulation. This is also true for the evolution of the albedo and the cloud coverage. 
This resilience of the runaway greenhouse is visible on the top panel of \fig{fig_OLR_reversed} through the hysteresis loop on the insolation. 

\subsection{Runaway greenhouse dry transition}
\label{sub_dry}

As explained in \sect{sect_methodology}, to study the transition of the runaway, we arbitrarily remove the surface ocean when the surface water vapor pressure reaches some threshold values (1 and 1.5\,bar). In the evaporation phase, we evaporate up to 2.6\,bar of water but we were not able to model the dry transition phase for vapor pressure higher than 1.5\,bar because of numerical instabilities. High vapor pressures induce a huge radiative imbalance leading to numerical failures similar to those discussed by \cite{kopparapu_habitable_2017}. 

\begin{figure}[!ht]
    \centering\includegraphics[width=\linewidth]{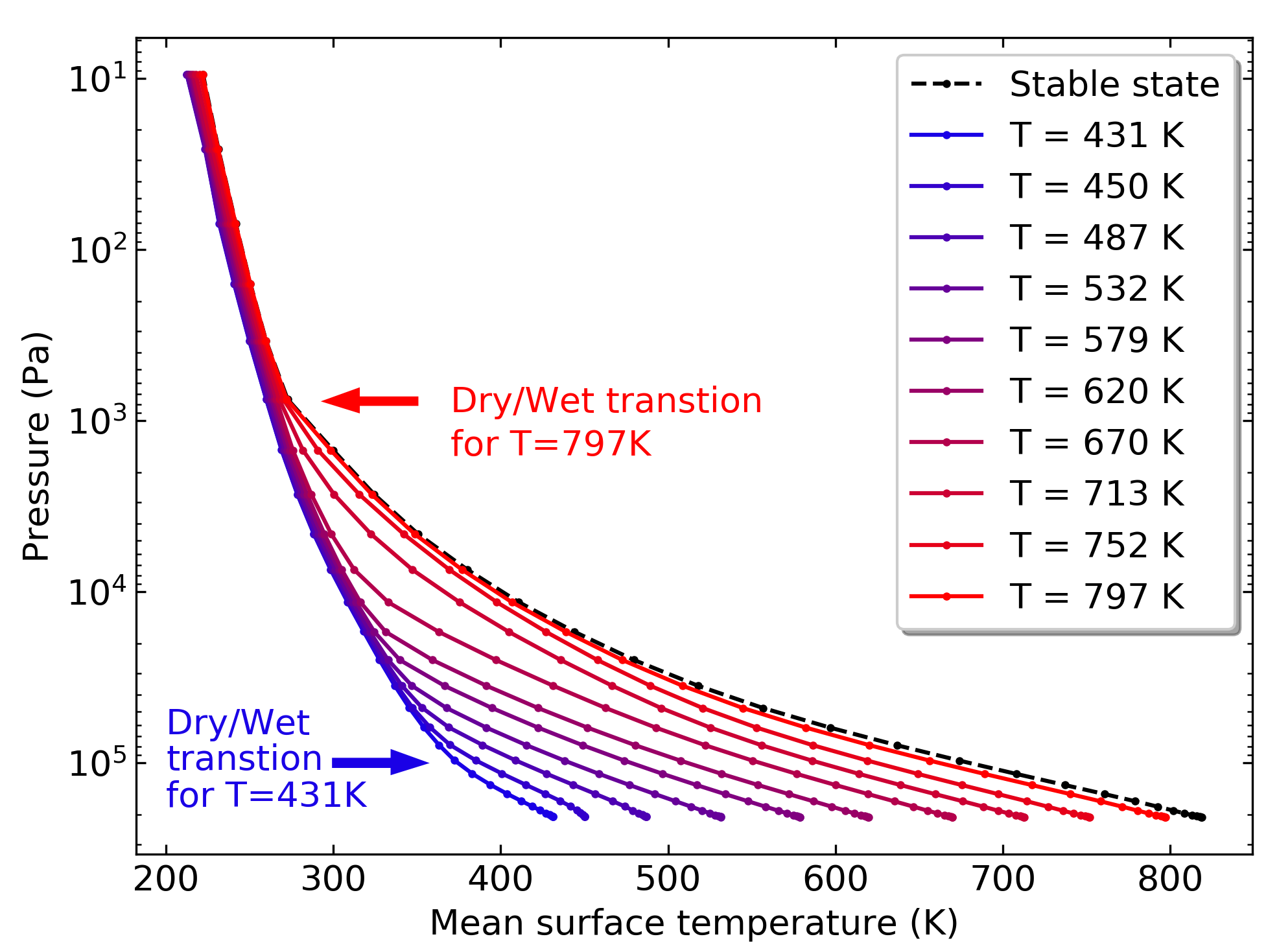}
    \caption{Temperature profile for different surface temperature during the dry transition with 1\,bar of water vapor. Arrows indicate the transition between a dry adiabat in the lower part of the atmosphere (high pressures) and a wet one in the bottom part (low pressures) for the coldest (blue) and hottest (red) cases. Dry region is below the transition while moist one is above \citep{abe_evolution_1988, kasting_runaway_1988}. The simulation setup is the waterworld with 1 bar of nitrogen without CO$_2$ (W1).}
    \label{fig_dry_PT_1bar}
\end{figure}

\begin{figure}[!ht]
    \centering\includegraphics[width=\linewidth]{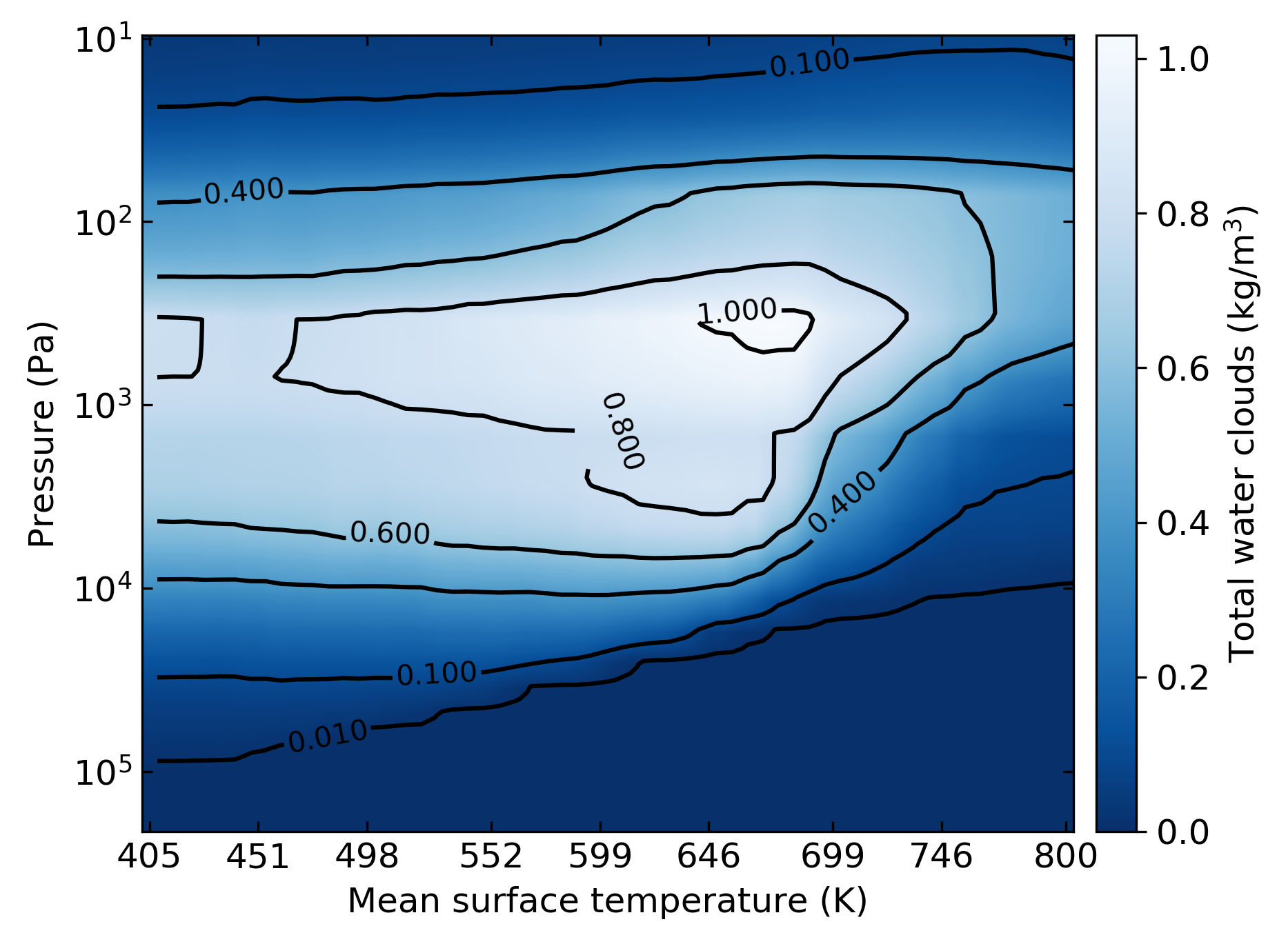}
    \caption{Map of the evolution of the mean vertical profile of the total water cloud (liquid + ice) content as a function of the global surface temperature during the dry transition phase. The simulation setup is the waterworld with 1 bar of nitrogen without CO$_2$ (W1)  assuming 1\,bar of water vapor. The profiles are averaged over one year.}
    \label{fig_Vmap_dry_1bar}
\end{figure}

\begin{figure*}[!ht]
    \centering\includegraphics[width=\linewidth]{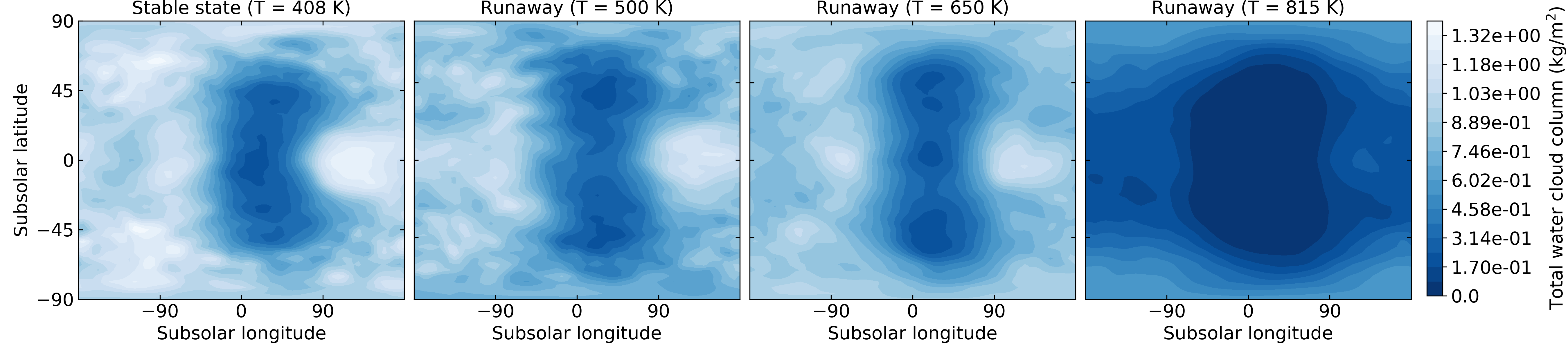}
    \caption{Latitude-longitude map of the cloud (liquid + ice) distribution during the dry transition phase with 1\,bar of vapor. The maps were calculated in the heliocentric frame, that is keeping the subsolar point at 0° longitude and 0° latitude, and using an annual average. The simulation setup is the waterworld assuming 1\,bar of nitrogen without CO$_2$.}
    \label{fig_latlon_clouds_dry}
\end{figure*}

\begin{figure}[!ht]
    \centering\includegraphics[width=\linewidth]{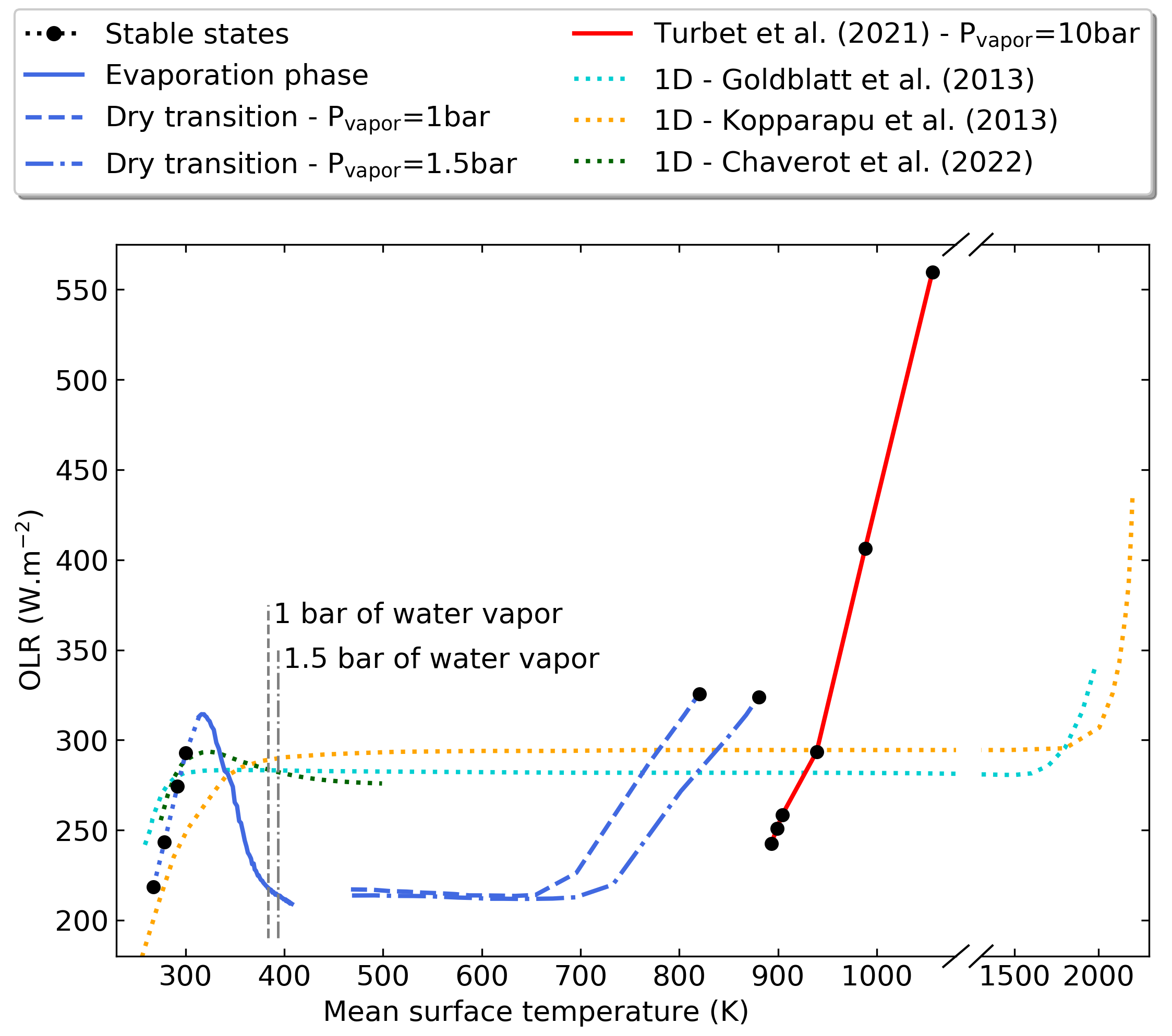}
    \caption{Evolution of the OLR as a function of the mean surface temperature from temperate stable states to post-runaway stable states. For blue curves, the simulation setup is the waterworld, 1\,bar of nitrogen without CO$_2$ (W1). The red curve is taken from \cite{turbet_day-night_2021} assuming 10\,bar of water vapor, 1\,bar of nitrogen and no CO$_2$. The insolation of the evaporation phase and the dry transition is equal to 400\,W.m$^{-2}$. The vertical grey lines are the temperatures for which 1\,bar and 1.5\,bar of water vapor are evaporated. Stable post-runaway states from \cite{turbet_day-night_2021} assume different insolations. The OLR values are averaged over 2 years which explain the gap between the evaporation and the dry transition phases. Cyan, orange and green dotted lines are OLR curves from reference works using 1D models and assuming 273\,bar of water (pure water from \citealt{goldblatt_low_2013}, Earth's case from \citealt{kopparapu_habitable_2013} and with 1\,bar of N$_2$ from \citealt{chaverot_how_2022} respectively).}
    \label{fig_dry_transition}
\end{figure}

We did two simulations for two different vapor surface pressures (1\,bar and 1.5\,bar corresponding to 10 and 15\,m of GEL) in order to explore the impact of the water reservoir on the obtained post-runaway states. 
We used the corresponding runaway greenhouse states as initial conditions then we remove the surface ocean from the simulation and we let the system evolve by keeping the atmospheric composition unchanged, as well as the insolation. 
As there is no humidity source anymore, the temperature profile of the atmosphere changes. 
More precisely, there is a transition from a previously wet to a dry adiabatic profile in the bottom part of the atmosphere (as seen in \citealt{boukrouche_beyond_2021} with a 1D model) as visible in \fig{fig_dry_PT_1bar}, thus we choose to name this second simulation phase the 'dry transition'. 
As the pressure/temperature profile of a dry adiabat is steeper than a wet one, the global surface temperature rises quickly. 
During the evaporation phase, the large thermal inertia of the ocean produces a cooling in the bottom part of the atmosphere preventing a quick increase of the surface temperature. 
In the dry transition, such forcing does not exist allowing a strong and fast increase of the global surface temperature.
For the evaporation phase (W1 simulation, ISR=400\,W.m$^{-2}$), more than a hundred years are needed to increase the temperature of about 80\,K (i.e. evaporating 1.5\,bar of water). While for the `dry' phase, only few tens of years are needed to increase the temperature of $\approx$400\,K.

Figure\,\ref{fig_Vmap_dry_1bar} shows the evolution of the total water cloud content (liquid + ice) during the dry transition. We see a re-evaporation of the bottom part of the cloud layer. This contribute to enrich the upper part of the atmosphere in water vapor. 
Moreover, due to the high temperatures at the surface, the water vapor migrates towards the upper layers and the vapor mass mixing ratio profile tends toward a constant value equal to 0.5\,kg.kg$^{-1}$ (see \app{app_vapor} for more details). 
We notice also a slightly increase of the cloud coverage at the very top of the atmosphere.

Figure\,\ref{fig_dry_transition} gives the evolution of the OLR as a function of the global mean surface temperature. The evaporation phase (below 420\,K) corresponds to the simulation W1 presented in \fig{fig_OLRvsT}. 
The blue dashed and dashed-dotted lines are respectively the dry transitions when 1\,bar and 1.5\,bar of water vapor is evaporated. The red curve is taken from \cite{turbet_day-night_2021} assuming 10\,bar of vapor, 1\,bar of nitrogen and no CO$_2$. 
Black dots are the different stables states obtained in each cases, and for different insolations. Blue and red curves are computed using different methods.
Blue curves are obtained by increasing the insolation from temperate stable states (warming simulation) while for the red curve the insolation is reduced step-by-step to find the tipping point for which condensation occurs (cooling simulation). 
We show that the size of the water reservoir determines the equilibrium temperature of the post-runaway state (black dots beyond 800\,K in \fig{fig_dry_transition}) as suggested by 1D models \citep{turbet_runaway_2019, boukrouche_beyond_2021, turbet_day-night_2021}. 
Dotted lines are reference OLR curves computed using different 1D models and for different atmospheric compositions. Cyan, orange and green dotted lines correspond to a pure water atmosphere \citep{goldblatt_low_2013}, the pre-industrial Earth \citep{kopparapu_habitable_2013}, and an atmosphere containing 1\,bar of nitrogen without CO$_2$ \citep{chaverot_how_2022} respectively. All these models consider a fully saturated atmosphere (see \sect{sub_rad_imbalance} for more details about differences between 3D and 1D saturated/sub-saturated models).

\begin{figure}[!ht]
    \centering\includegraphics[width=\linewidth]{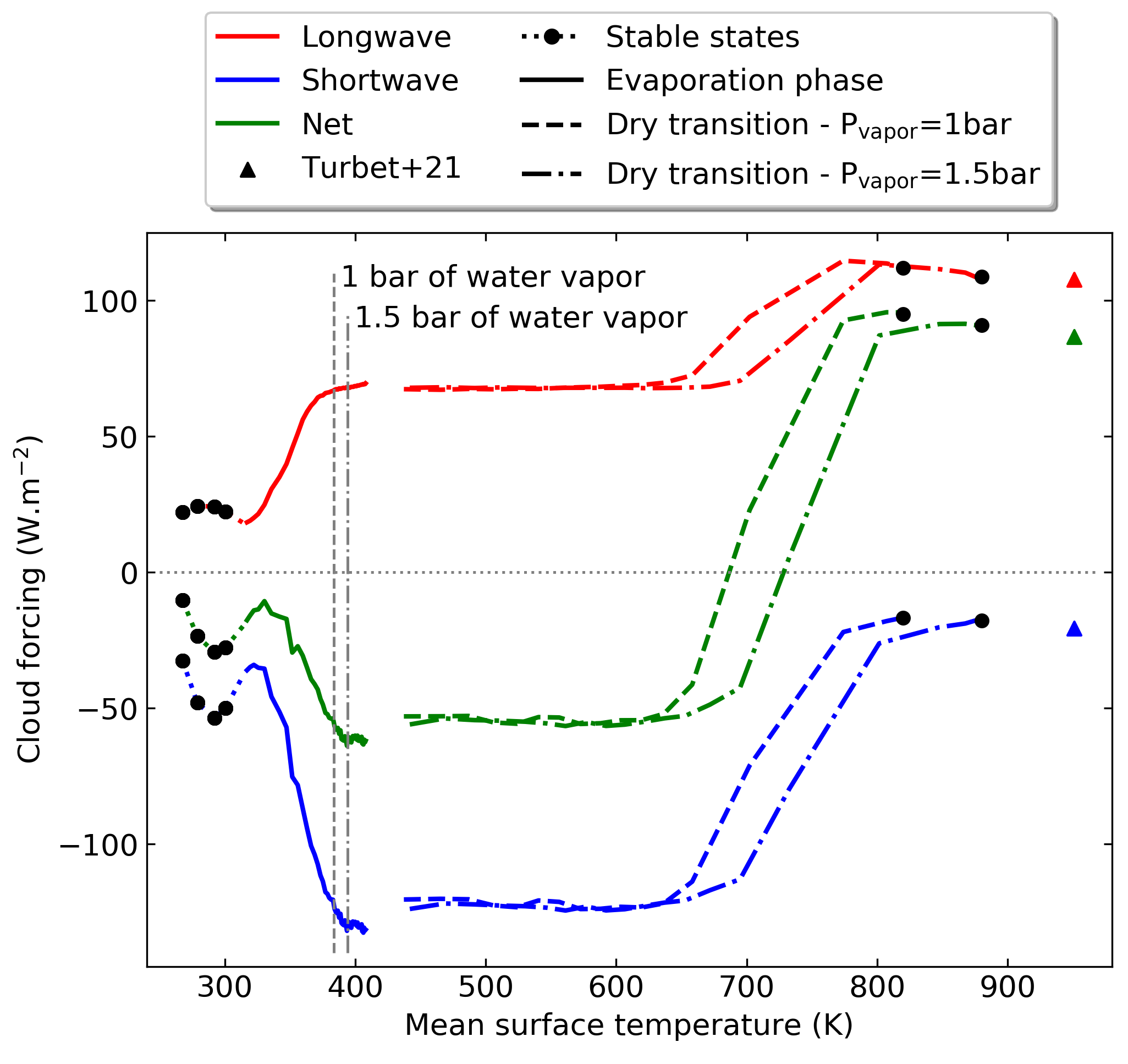}
    \caption{Evolution of the cloud forcing during the runaway greenhouse for a waterworld with 1 bar of nitrogen without CO$_2$ (W1). Solid lines are the evaporation and dry transition phases while black dots are stable states assuming a 2-years average. Negatives values indicate a cooling effect of the clouds and positive ones a warming effect.
    In the longwave (blue curves), clouds have a cooling effect due to their high albedo reflecting  the incoming stellar flux back to space. The red curve correspond to the warming contribution due to the greenhouse effect of the clouds (i.e. thermal mission absorbed by the clouds). The green curve is the net effect of the clouds, including shortwave and longwave contributions. The vertical grey lines are the temperatures for which 1\,bar and 1.5\,bar of water vapor are evaporated. There are used as initial conditions for the dry transition phase (colored dotted lines). Colored triangles are the values obtained in \cite{turbet_day-night_2021} for the same insolation (400\,W.m$^{-2}$) but including 10\,bar of water vapor. The fluxes values are averaged over 2 years which explain the gap between the evaporation and the dry transition phases}
    \label{fig_cloud_forcing}
\end{figure}

\begin{figure}[!ht]
    \centering\includegraphics[width=\linewidth]{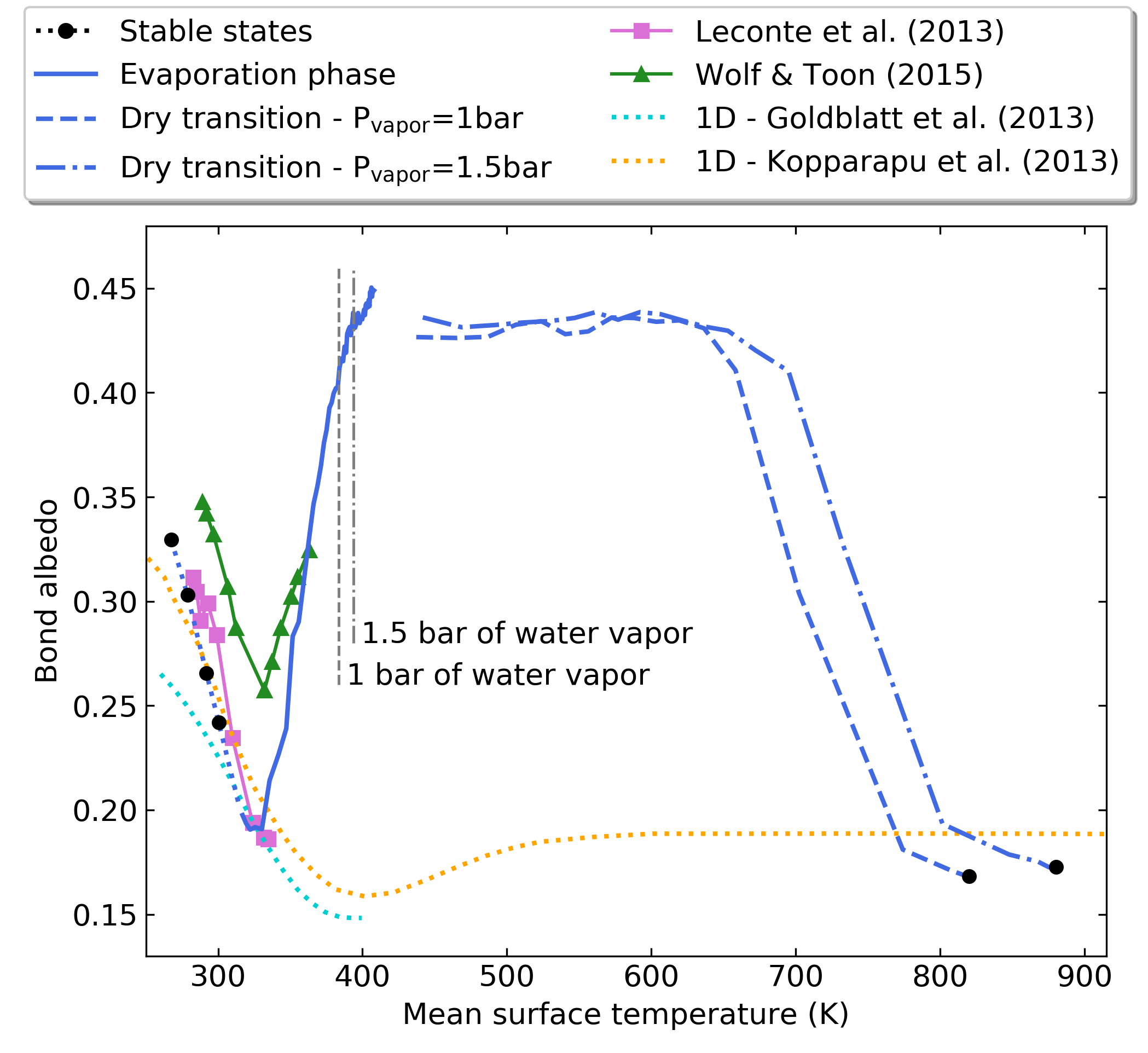}
    \caption{Evolution of the bond albedo as a function of the mean surface temperature from temperate stable states to post-runaway stable states. The simulation setup is the waterworld with 1 bar of nitrogen without CO$_2$ (W1). The insolation of the evaporation phase and the dry transition is equal to 400\,W.m$^{-2}$. The vertical grey lines are the temperatures for which 1\,bar and 1.5\,bar of water vapor are evaporated. The albedo values are averaged over 2 years which explain the gap between the evaporation and the dry transition phases. Pink squares and green triangles are steady states of the Earth's case from \cite{leconte_increased_2013} and \cite{wolf_evolution_2015} respectively. Dotted lines are albedo curves from reference works using 1D models (1\,bar of N$_2$ from \citealt{goldblatt_low_2013} and Earth's case from \citealt{kopparapu_habitable_2013}). }
    \label{fig_albedoT_all}
\end{figure}

The reason for which we consider small amounts of water (1 and 1.5\,bar) come from the fact that the addition of vapor in the atmosphere increases its thermal inertia thus decreasing the evaporation rate. 
Due to numerical constrains we were not able to evaporate more than 2.6\,bar of water vapor in our simulations. Moreover, we were not able to model the dry transition phase itself for vapor pressures beyond 1.5\,bar because of numerical instabilities. 
Indeed, for such pressures the radiative unbalance skyrockets inducing numerical failures similar to those described by \cite{kopparapu_habitable_2017} for the onset of the runaway greenhouse.
There is no strong evolution of the wind pattern during the dry transition. 
The day-to-night cloud dichotomy discussed in \sect{sub_cloud_coverage} remains during this phase as shown by \fig{fig_latlon_clouds_dry} and becomes even more contrasted.
It is accentuated by the re-evaporation of the clouds to converge on the exact same pattern as described in \cite{turbet_day-night_2021} with a purely cloud free day-side.
OLR values obtained in this work are also consistent with those obtained for different insolations in \cite{turbet_day-night_2021}. 
Moreover, even with a quantity as high as 10\,bar of water vapor in the atmosphere, they obtain global circulation, temperature and cloud distributions strongly similar to those presented in this section. Therefore, we can reasonably expect the same climate pattern and evolution for post-runaway states with higher water pressures (few tens of bar).

Figure\,\ref{fig_cloud_forcing} shows the evolution of the cloud forcing during both phases of the runaway greenhouse. For temperate habitable states, the cooling effect of the clouds in the shortwave (blue curve) is more efficient than the warming due to their greenhouse effect in the longwave (red curve).
During the evaporation phase, as the cloud deck densifies, the Bond albedo increases (\fig{fig_albedoT_all}) and the colling effect of the clouds is even more efficient (blue curve in \fig{fig_cloud_forcing}).
However, when the surface ocean is considered entirely evaporated (vertical grey dotted lines at 383\,K and 393\,K corresponding respectively to 1\,bar and 1.5\,bar), after a strong increase of the global temperature the net radiative effect of the clouds switch from cooling to warming (green curve) around 700\,K.
We show that during the dry transition phase, as the Bond albedo decreases (see \fig{fig_albedoT_all} and blue curve in \fig{fig_cloud_forcing}), the greenhouse effect of clouds becomes the dominant process and the net effect of the clouds tends to warm the planet (\fig{fig_cloud_forcing}). In other words, the thick layer of clouds at the top of the atmosphere acts as a shield that retains the thermal emission of the planet.
The cloud forcing of the obtained post-runaway states is very similar to the one from \cite{turbet_day-night_2021} even if they consider 10 times more water vapor. 
This suggest that 1\,bar of vapor is enough to stabilize the climate in a post-runaway configuration, with a thick cloud deck at the top of the atmosphere, preferentially on the night side of the planet.
The temperature for which the net effect of the clouds switch from cooling to warming is concomitant with the increase of OLR and function of the water vapor content. More vapor and clouds will require a higher temperature to evaporate and modify the cloud albedo.
The 50\,K gap between the evaporation phase and the dry transition on \figss{fig_dry_transition}{fig_cloud_forcing}{fig_albedoT_all} is due to the very fast evolution of the system during the first stages of the simulation.

\section{Discussion}
\label{sect_discussion}

\subsection{Literature comparison}

\begin{figure}[!ht]
    \centering\includegraphics[width=\linewidth]{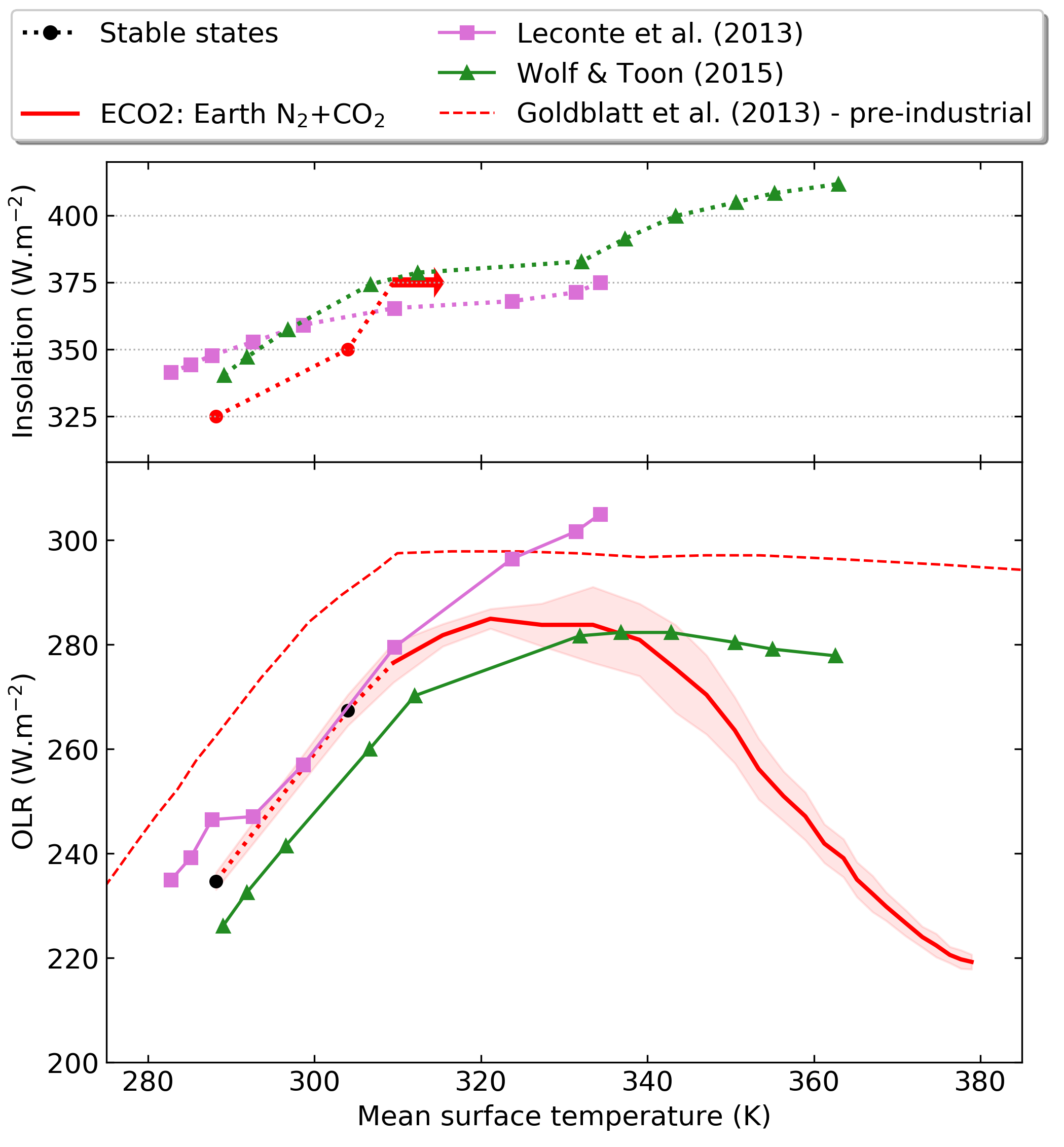}
    \caption{Literature comparison of the OLR and the insolation as a function of the temperature. The black dots are the stables states corresponding to various insolations while the full curves are the unstable state of the runaway greenhouse.  The top panel indicates the insolation values of each stable state. The colored arrows represent the insolation for which the runaway greenhouse arises for each setup. The OLR and temperature values are averaged over 2 years and the colored fills are the 1-sigma uncertainties due to variability on the OLR calculation. Pink squares and green triangles are steady states of the Earth's case from \cite{leconte_increased_2013} and \cite{wolf_evolution_2015} respectively. The red dashed curve is a reference result from \cite{goldblatt_low_2013} for a pre-industrial atmosphere and using a 1D model assuming saturation mixing ratio below 5\%.}
    \label{fig_literature}
\end{figure}

We show in \sect{sub_rad_imbalance} that the radiative imbalance is constant during all the evaporation phase.
Similarly to \cite{leconte_increased_2013}, we do not find any moist stable states while \cite{wolf_evolution_2015}, using CAM4, or \cite{popp_transition_2016}, using ECHAM6, do.
Multiple discussions in the literature about this still open question \cite[e.g.][]{kopparapu_habitable_2017,yang_simulations_2019} pointed out that the specific humidity of the stratosphere is a key factor to explain this difference. 
Indeed, the GCM inter-comparison done by \cite{yang_simulations_2019} highlights that the Generic-PCM\footnote{named LMDG in \cite{yang_simulations_2019}} tends to be wetter than CAM4 for planets around G-type stars. 
In the same time, the cloud fraction and the cloud thickness is higher in CAM4 inducing a cooling of the planet.
This was discussed in \cite{kopparapu_habitable_2017}, showing that \cite{wolf_evolution_2015} found higher stratospheric temperatures than \cite{leconte_increased_2013} for a same surface temperature. 
High relative humidity in the stratosphere reduces the thermal emission from the upper layers, allowing a larger difference between the OLR and the ASR. This is the commonly accepted answer to explain the existence - or not - of a moist stable state.
More inter-comparisons based on the method used in \cite{yang_simulations_2019}, as well as high-resolution cloud resolving simulations \citep[e.g.][]{lefevre_3d_2021}, should be done to solve this question.
By using a 1D climate model, \cite{popp_initiation_2015} suggests that the convection scheme could be the answer to explain potential moist runaway greenhouse. The number of Cloud Condensation Nucleii (CCN) may also impact the cloud formation thus the possibility of a moist state.

We compare our results for the Earth case with pre-industrial CO$_2$ quantities (ECO2) with major results from the literature \citep{leconte_increased_2013,wolf_evolution_2015} in \fig{fig_literature}. We do not show a direct comparison of the ASR but we compare the Bond albedo from simulation W1 with \cite{leconte_increased_2013} and \cite{wolf_evolution_2015} in \fig{fig_albedoT_all}. 
As shown by \fig{fig_literature}, we found slightly different global surface temperature and OLR values compared to \cite{leconte_increased_2013} and \cite{wolf_evolution_2015}, even for a same insolation. 
This is not surprising regarding numerous differences between the models highlighted by \cite{yang_simulations_2019}. 
In \cite{leconte_increased_2013}, they used the Generic-PCM as this work but they mimic the ocean thank to a large thermal inertia of the ground while we use a more complex slab ocean (without heat transport). 
This, combined with the development of the model, may explain the differences between our results and especially on the relation between insolation and the global surface temperature. 
Even if \cite{wolf_evolution_2015} found a moist stable state, the evolution of the OLR as a function of the global surface temperature is qualitatively consistent with our simulations.

The red dashed line is taken from \cite{goldblatt_low_2013} and correspond to a pre-industrial composition with a subsatured atmosphere. This curve is comparable with our ECO2 case to estimate the impact of 3D processes. The constant offset of OLR values for temperate states between 1D and 3D simulations is due to the greenhouse effect of the clouds. In the same way, the large difference of OLR during the runaway greenhouse is due to the contribution of the thick cloud coverage. As discussed several works \citep[e.g.][]{ishiwatari_numerical_2002,abe_habitable_2011,leconte_increased_2013}, reducing the relative humidity of 1D models can produce results in a good agreement with 3D modelling for temperate stable states. However, large differences remain for the runaway greenhouse as the key process is the evolution of the cloud coverage not estimable using 1D models. 
From a qualitative point of view, the shape of the OLR curve we show is similar to historical estimations of the thermal emission of an atmosphere composed of H$_2$O and N$_2$ \citep[e.g.][]{nakajima_study_1992}, but the quantitative values are different because of the addition of more complex physical processes (non-grey atmosphere, clouds, global dynamics).

\subsection{Limitations and prospects}
The H$_2$O+N$_2$ correlated-k table used in this work was made by \cite{chaverot_how_2022} using the HITRAN database \citep{gordon_hitran2016_2017}. \cite{goldblatt_low_2013} showed that the HITEMP database is more accurate and complete than HITRAN for high temperatures and this impacts the estimation of OLRs and ASRs (see also \citealt{kopparapu_habitable_2013} and \citealt{ramirez_terrestrial_2014}).
More precisely, they highlight using a 1D climate model, that HITRAN overestimates the albedo and thus underestimates the ASR for temperatures beyond $\approx$360\,K. 
As in 3D GCM simulations the relative humidity (thus the vapor partial pressure) is lower than in 1D models \citep{leconte_increased_2013}, we can assume no significant difference in our simulations up to $\approx$400\,K.
Therefore, this does not affect much the evaporation phase but may change the dry transition phase and the final equilibrium temperature by slightly modifying the Bond albedo of the planet an the heating rates. 
This could be investigated but we are confident on the global evolution described in this paper and on the reliability of general mechanisms which determine the climate pattern transition between pre- and post-runaway. We describe general processes and tendencies and our results are in accordance with previous works using different GCMs.

In every simulation, the heat capacity (c$_{\rm p}$) and the mean molecular weight are assumed to be constant. This is inaccurate when the temperature increases strongly (because of the dependence of c$_{\rm p}$ on the temperature) and when the dominant gas changes (from nitrogen dominated to water dominated during the evaporation phase, see \fig{fig_Vprofile_clouds} and \citealt{chaverot_how_2022}). 
Our version of the Generic-PCM does not include variable heat capacity and molecular weight but ongoing developments will allow to study the impact of variations of these quantities in a near future. For consistency reasons, we do not update the value of c$_{\rm p}$ when we initialize the simulation of the dry transition phase. 

In this work, we describe in detail the simulation with 1\,bar of nitrogen without CO$_2$ (W1) to explore the effect of the water without additional absorption sources (nitrogen is assumed to be optically thin, except for the CIA). 
Therefore, the minimal absorption given by the correlated-k table is for a water volume mixing ratio equal to 10$^{-6}$. 
When the mixing ratio is lower, for example in the stratosphere at low temperature, the water vapor absorption is given by this threshold value.  
This approximation may induce an error for very low water content but becomes negligible when water is the dominant gas, in other words during the runaway greenhouse on which this work is focused on. 

In \sect{sub_dry}, we describe a transition between a wet to a dry adiabat. This is true for low partial water pressures but as shown by \cite{selsis_cool_2023} it becomes wrong for pressures higher than roughly 1\,bar (function of the considered host star). In this case, the temperature profile follows a radiative profile. 

Finally, as we do not search for the exact tipping point triggering the runaway greenhouse we probably overestimate the radiative imbalance by a few W.m$^{-2}$. This does not change the physics described in this work, but a study focused on this could be interesting in order to estimate the evaporation timescale of different water reservoirs.

\section{Conclusions}

We study the runaway greenhouse with the 3D Generic-PCM in order to link temperate stables states to post-runaway stable states. 
The positive feedback of the runaway greenhouse is triggered by the strong absorption of the water vapor and driven by a large water inventory (i.e. surface ocean) available through evaporation.
From terrestrial climate conditions, we increase the insolation step-by-step in order to find the onset of the runaway greenhouse, following the method used by previous works \citep{leconte_increased_2013, wolf_delayed_2014, popp_transition_2016}.
The originality of this work is to go beyond the onset and to study the transition toward post-runaway states. We show that there is two phases: 
\begin{enumerate}
    \item An "evaporation phase", for which the water vapor content and the cloud thickness increases rapidly, 
    \item When the ocean is considered entirely evaporated, a "dry transition phase" characterized by a switch from a wet adiabatic profile to a dry adiabatic profile. During this second step, the temperature increases quickly, then the climate converges onto a new stable state at high temperature: a post-runaway state. 
\end{enumerate}

We describe the evolution of the composition of the atmosphere as well as the changes of the cloud pattern in \sect{sub_evap}. We show that even if the tipping point of the runaway greenhouse depends on the simulation setup (e.g. continents, CO$_2$ as a minor constituent), the evolution path during the transition does not, for a given background gas pressure. 
As the bottom part of the atmosphere is water dominated, it becomes optically thick and the physics is decoupled from the surface conditions. 
In the same way, a low content of CO$_2$ (376\,ppm) does not influence the climate beyond the onset (\fig{fig_OLRvsT}).

The main characteristic of the first phase of the runaway greenhouse (evaporation phase) is the evolution of the cloud coverage. It grows because of the intense evaporation and it migrates toward the upper layers (\fig{fig_Vprofile_clouds}). 
This migration increases the Bond albedo creating a drop of the ASR (\fig{fig_OLR_OLRcsvsT}). The day to night distribution of the clouds is also affected. As shown by \fig{fig_latlon_clouds}, the night-side is more cloudy than the day-side which tend to increase the greenhouse effect of the clouds. This competition between the warming and cooling effect of the clouds is shown in \fig{fig_albedoT_all}. We highlight that during the entire evaporation phase, the net effect of the clouds tends to cool down the planet. 
However, the addition of water vapor due to evaporation creates an extra absorption able to overshoot this cooling, leading to a roughly constant radiative imbalance (\fig{fig_OLR_OLRcsvsT}) which warms continuously the planet. 
The general atmospheric circulation is strongly affected by the evaporation and the evolution of the cloud coverage. As the top part of the atmosphere becomes wetter, it absorbs more stellar flux and the heating rate rises creating an intense global circulation \citep{fujii_nir-driven_2017}.
As visible in \fig{fig_wind}, strong stratospheric jets appear substituting the usual Hadley cells of the temperate Earth. 
We explore also the potential reversibility of the runaway greenhouse. The idea is to use a runaway state as a initial state, then to reduce the insolation in order to condense back the water in the surface ocean. We show in \fig{fig_OLR_reversed} that the radiative unbalance characteristic of the runaway greenhouse creates a hysteresis loop on the insolation making it resilient to any small variations of incoming flux.
We also show that if a re-condensation happens, the OLR increases when the surface temperature decreases, following the exact same evolution pathway than during the evaporation phase.

In a second step, when a given quantity of water is evaporated (1\,bar or 1.5\,bar), we consider the ocean as entirely evaporated and we remove it from the simulation (see \sect{sub_dry}). 
The absence of humidity sources at the bottom of the atmosphere induced a transition from a wet to a dry adiabatic profile (\fig{fig_dry_PT_1bar}). Water vapor migrates towards the upper layers of the atmosphere and lower clouds evaporate (\fig{fig_Vprofiles_dry_1bar}). 
We observed also a more contrasted dichotomy between a cloudy night-side and a cloud free day-side \fig{fig_latlon_clouds_dry}, which is consistent with results from \cite{turbet_day-night_2021} studying post-runaway states for similar atmospheres. 
Because of the re-evaporation of the clouds, the albedo decreases (\fig{fig_albedoT_all}) and the their net effect tend to warm the planet (\fig{fig_cloud_forcing}, see also \citealt{turbet_day-night_2021}).  
This reduces also the radiative unbalance, allowing the simulation to converge on a post-runaway state as shown in \fig{fig_dry_transition}. This final state strongly depends on the quantity of water available in the atmosphere, in other words on the size of the initial water reservoir. 

Finally, by comparing our results to historical studies using 1D models, we show that 3D processes (clouds and dynamics) impact strongly the evolution of the climate during the first stages of the runaway greenhouse as well as during the dry transition phase. This is of major importance to discuss potential observability of this kind of planet with reflected or emitted light.

\section*{Acknowledgements}

This work has been carried out within the framework of the NCCR PlanetS supported by the Swiss National Science Foundation under grants 51NF40\_182901 and 51NF40\_205606. 
GC and EB acknowledge the financial support of the SNSF (grant number: 200021\_197176 and 200020\_215760)
MT thanks the Gruber Foundation for its generous support to this research. MT acknowledges support from the Tremplin 2022 program of the Faculty of Science and Engineering of Sorbonne University. 
The authors thank the Generic-PCM team for the teamwork development and improvement of the model. 
The computations were performed at University of Geneva on the Baobab and Yggdrasil clusters. 
This research has made use of NASA's Astrophysics Data System.
We thank the two referees Ravi K. Kopparapu and Colin Goldblatt for their helpful comments.

\section*{Data availability}

The GCM outputs of the stable states (temperate and post-runaway) of each setup, as well as several snapshots along the runaway transition for setup W1 are available here: \url{https://zenodo.org/record/8325631}. Other data underlying this article will be shared on request to the corresponding author.
The Generic-PCM (and documentation on how to use the model) can be downloaded from the SVN repository \url{https://svn.lmd.jussieu.fr/Planeto/trunk/LMDZ.GENERIC/} (version 2521). More information and documentation are available on \url{http://www-planets.lmd.jussieu.fr}.

\bibliographystyle{aa}
\bibliography{biblio}

\appendix
\section{Evolution of water vapor and clouds during the runaway greenhouse}
\label{app_vapor}

\begin{figure}[!ht]
    \centering\includegraphics[width=\linewidth]{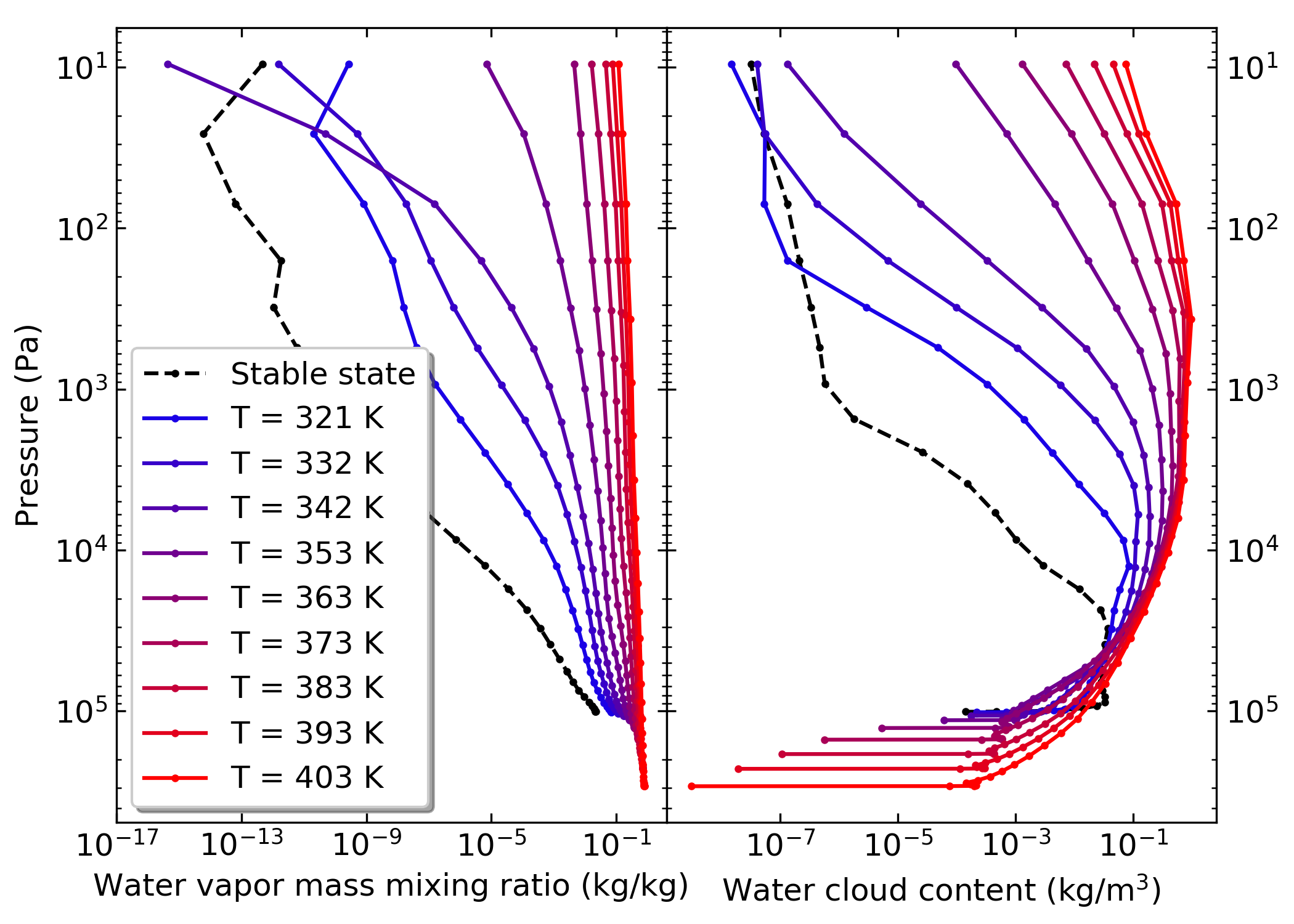}
    \caption{Evolution of the water vapor mass mixing ratio  and of the mean vertical profile of the water cloud (liquid + ice) content during the evaporation phase. The simulation setup is the waterworld with 1 bar of nitrogen without CO$_2$ (W1). The profiles are averaged over one year. The dashed black line corresponds to the temperate stable state with the highest insolation (ISR=375\,W.m$^{-2}$). }
    \label{fig_Vprofile_clouds}
\end{figure}

As discussed in \sect{sect_results}, the two phases of the runaway greenhouse strongly modify the distribution of water vapor as well as the cloud coverage. In this section, we provide additional information concerning these quantities. 
During the evaporation phase, the evolution of the average water vapor profile (left panel of \fig{fig_Vprofile_clouds}) shows a strong enrichment of the upper layers of the atmosphere in water vapor from 1$\times10^{-13}$\,kg.kg$^{-1}$ to 0.1\,kg.kg$^{-1}$. This is coherent with the densification and the drift of the cloud coverage towards the stratosphere shown on the right panel of \fig{fig_Vprofile_clouds} (this panel corresponds to the results presented in \fig{fig_Vmap_clouds}). This drift of the clouds is also discussed in \cite{kopparapu_habitable_2017}. 

\begin{figure}[!ht]
    \centering\includegraphics[width=\linewidth]{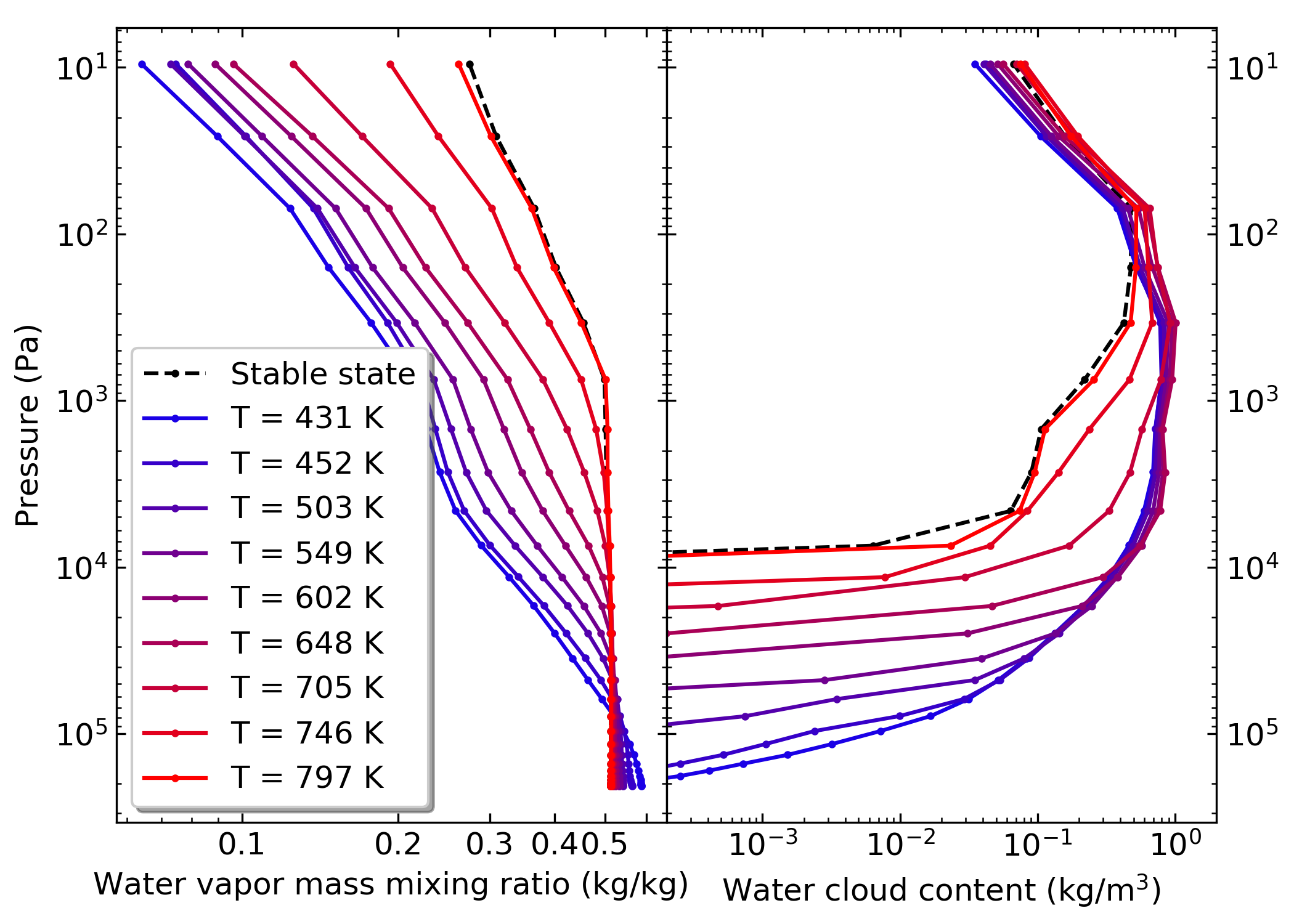}
    \caption{Evolution of the water vapor mass mixing ratio and of the mean vertical profile of the water cloud (liquid + ice) content during the dry transition phase. The simulation setup is the waterworld with 1 bar of nitrogen without CO$_2$ (W1) assuming 1\,bar of water vapor. The dashed black line corresponds to the post-runaway stable state obtained after the runway greenhouse transition.}
    \label{fig_Vprofiles_dry_1bar}
\end{figure}

During the dry transition, as there is no moisture source from the surface anymore, the surface temperature rises, leading to a migration of the water vapor and of the water clouds toward the upper layers of the atmosphere.
The left panel of \fig{fig_Vprofiles_dry_1bar} shows the evolution of the water vapor mass mixing ratio profile. We see that the bottom part of the atmosphere become dryer (i.e. the water vapor content decreases) and the mass mixing ratio profile tends toward a constant value up to 1$\times10^{3}$\,Pa. This is the pressure above which the pressure temperature conditions are favorable to form clouds (see left panel of \fig{fig_Vprofiles_dry_1bar} and \fig{fig_Vmap_dry_1bar}).  
In the same time, the bottom part of the cloud layer evaporate due to the increase of surface temperature. This is clearly visible on the right panel in \fig{fig_Vprofiles_dry_1bar} (this panel corresponds to the results presented in \fig{fig_Vmap_dry_1bar}).

\begin{figure}[!ht]
    \begin{subfigure}[b]{\linewidth}
        \centering\includegraphics[width=\linewidth]{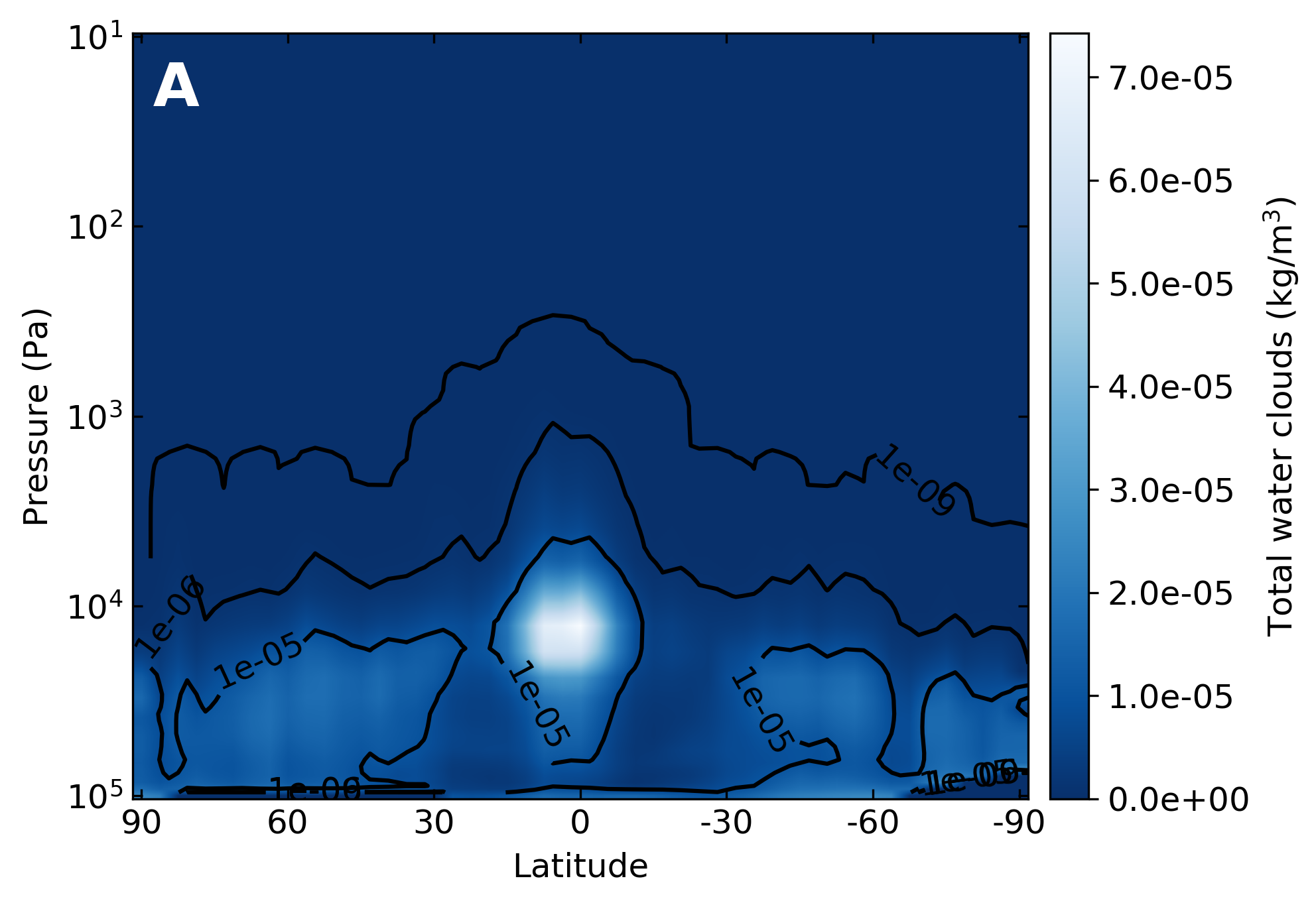}
    \end{subfigure}\\
    \begin{subfigure}[b]{\linewidth}
        \centering\includegraphics[width=\linewidth]{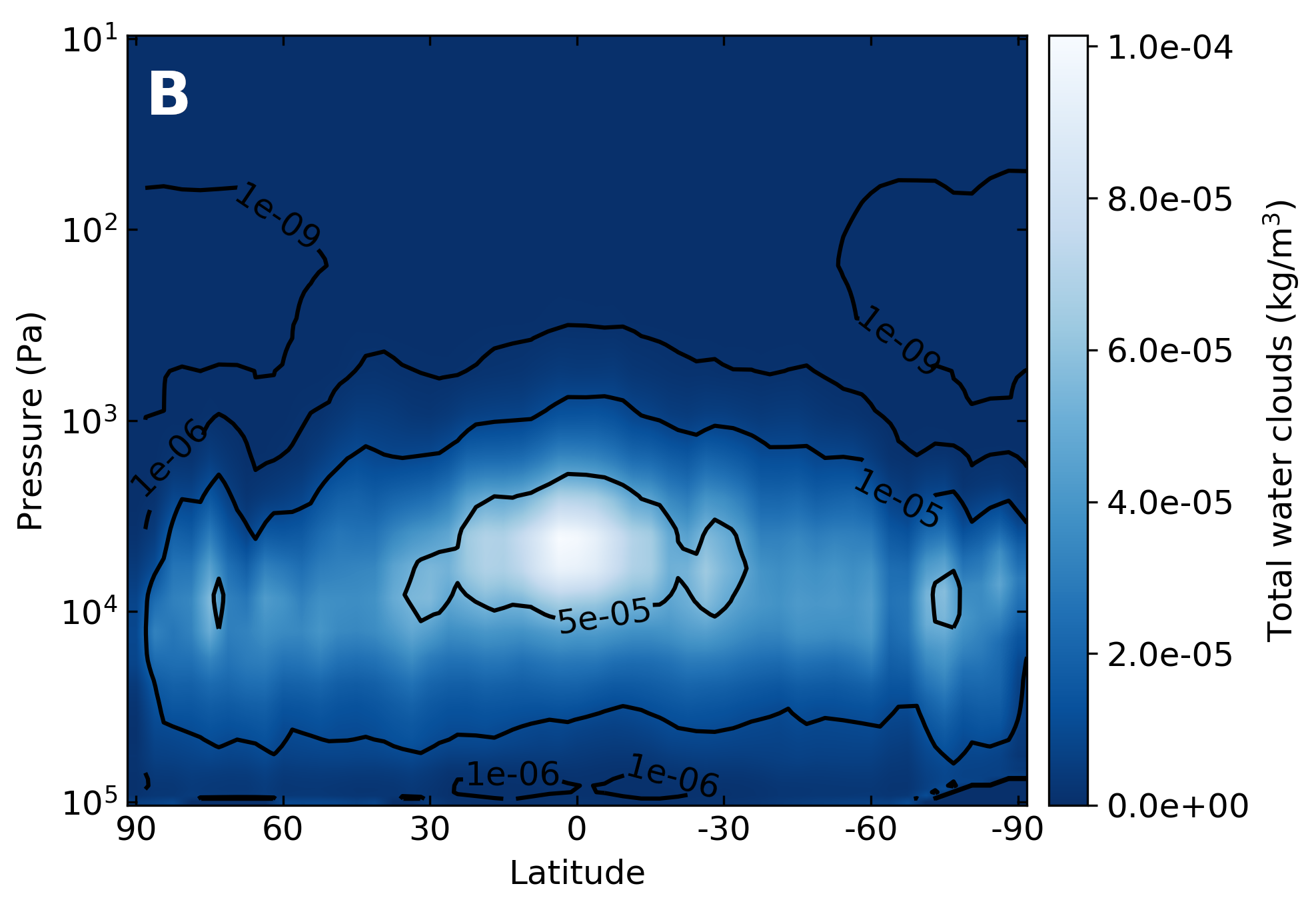}
    \end{subfigure}\\
    \begin{subfigure}[b]{\linewidth}
        \centering\includegraphics[width=\linewidth]{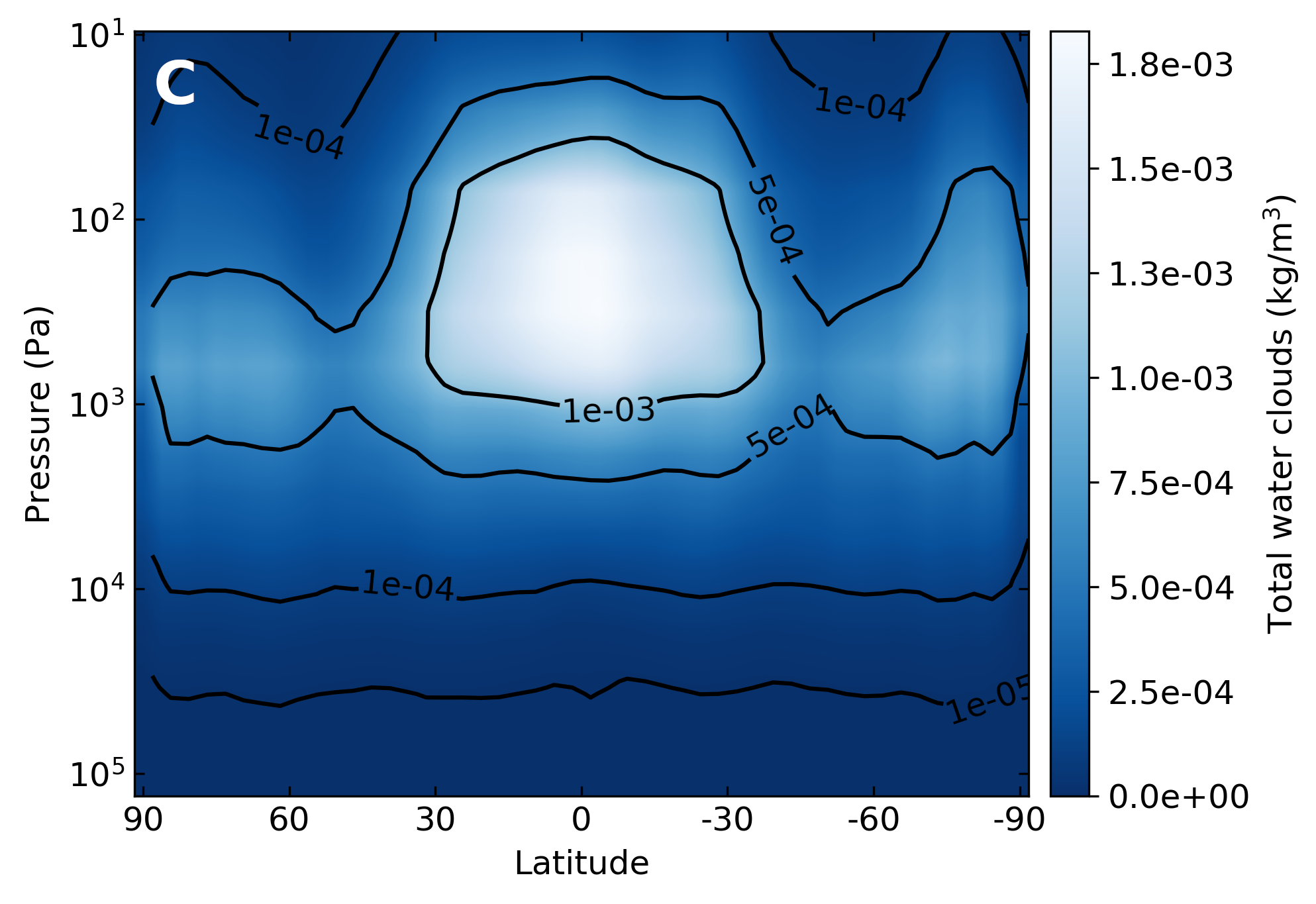}
    \end{subfigure}
    \caption{Evolution of the latitude-altitude map of the water cloud (liquid + ice) distribution during the evaporation phase. The simulation setup is the Earth with 1 bar of nitrogen without CO$_2$ (E1). The maps are 2-years averaged. The A, B and C panels correspond to global surface temperatures of 320\,K, 340\,K and 370\,K respectively.}
    \label{fig_lat-lon_clouds}
\end{figure}

The latitudinal distribution of the clouds coverage during the runaway greenhouse is presented in \fig{fig_lat-lon_clouds} for global surface temperatures of 320\,K, 340\,K and 370\,K (panels A, B and C respectively). The setup is the Earth without CO$_2$ (E1).  While the usual cloud pattern we see on the Earth remains during the onset of the runaway greenhouse (panel A), it is modified when the water vapor content becomes  sufficient to drift the cloud toward the upper atmosphere (panel B). 
When the stratospheric circulation described in \sect{sub_winds} appears (panel C which corresponds to a global surface temperature equal to 370\,K), the cloud deck is located at the top of the atmosphere with a maximum density around the equator where the upward winds are the strongest. Cloud minima are around the tropics because of the downward winds (see \fig{fig_wind} for the wind circulation).

\subsection{Effect of the out-of-equilibrium evolution}

\begin{figure}[!ht]
    \centering\includegraphics[width=\linewidth]{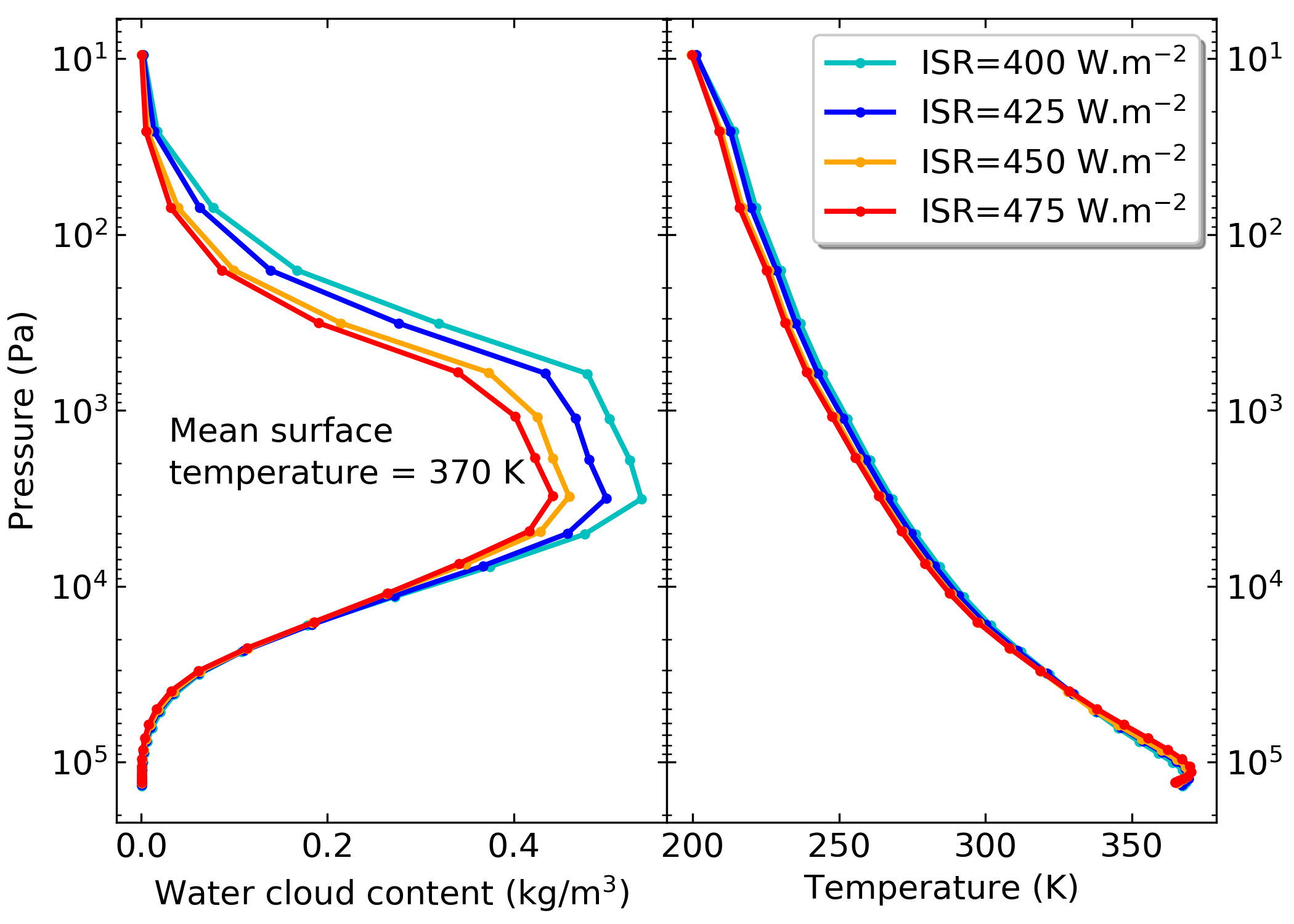}
    \caption{Cloud (liquid + ice) profiles and temperature profiles during the runaway greenhouse when the global surface temperature is equal to 370\,K, from simulations assuming different insolations and a common initial condition. The simulation setup is the waterworld with 1 bar of nitrogen without CO$_2$ (W1). The profiles are averaged over one year.}
    \label{fig_Vprofile_clouds_insolation}
\end{figure}

By definition, the runaway greenhouse is an out-of-balance process making its analysis quite complex. In order to study the impact of the insolation on the evolution of the climate, we performed four different simulations with the setup W1 and various insolations but sharing the same initial conditions. 
Results are shown in \fig{fig_Vprofile_clouds_insolation}, where the left panel is the vertical water cloud profile, and the right panel is the temperature profile (both annually averaged). For every curve the global surface temperature is equal to 370\,K. 
For high insolations, the cloud coverage is less dense at equivalent global surface temperature (left panel), and where the temperature is lower in the upper layers (right panel). Due to highly imbalanced radiative conditions at high insolations, the evolution is so fast that there is a delay of the growth of the cloud coverage compared to the increase of global surface temperature.  
This affects the global evolution of the atmosphere by reducing the cooling effect of the clouds at a given surface condition. The enrichment in water vapor of the top part of the atmosphere is also delayed. However, as the cloud coverage is less dense at high insolations, the difference between cloud maxima at the equator and cloud minima around the tropics is less contrasted, smoothing the latitudinal distribution of both the water vapor and the temperature.

\end{document}